\documentclass[10pt,journal,compsoc]{IEEEtran}
\usepackage{graphicx}
\usepackage{color}
\usepackage{amssymb,amsmath}
\usepackage{balance}  
\usepackage{algorithm}
\usepackage{subfigure}
\newtheorem{problem} {Problem}
\newtheorem{lemma} {Lemma}
\newtheorem{theorem} {Theorem}
\newtheorem{example}{Example}
\usepackage{hyperref}
\usepackage{url}
\DeclareMathOperator*{\argmin}{arg\,min}

\ifCLASSINFOpdf
\else
\fi

\begin{document}
\title{Scalable Temporal Latent Space Inference for Link Prediction in Dynamic Social Networks}
\author{Linhong Zhu,
        Dong Guo,
        Junming Yin,
        Greg Ver Steeg,
        Aram Galstyan\\
\thanks{Linhong Zhu, Greg Ver Steeg and Aram Galstyan are with Information Sciences Institute, University of Southern California, email: \{linhong, gregv, galstyan\}@isi.edu.}
\thanks{Dong Guo is with the Department of Computer Science, University of Southern California, email: dongguo@usc.edu.}
\thanks{Junming Yin is with the Department of Management Information Systems, University of Arizona, email: junmingy@email.arizona.edu.}}
\IEEEcompsoctitleabstractindextext{
\begin{abstract}
We propose a temporal latent space model for link prediction in dynamic social networks, where the goal is to predict links over time based on a sequence of previous graph snapshots. The model assumes that each user lies in an unobserved latent space, and interactions are more likely to occur between similar users in the latent space representation. In addition, the model allows each user to gradually move its position in the latent space as the network structure evolves over time. We present a global optimization algorithm to effectively infer the temporal latent space. Two alternative optimization algorithms with local and incremental updates are also proposed, allowing the model to scale to larger networks without compromising prediction accuracy. Empirically, we demonstrate that our model, when evaluated on a number of real-world dynamic networks, significantly outperforms existing approaches for temporal link prediction in terms of both scalability and predictive power.
\end{abstract}
}

\maketitle
\section{Introduction}\label{sec:intro}
Understanding and characterizing the processes driving social interactions is one of the fundamental problems in social network research. A particular instance of this problem, known as {\em link prediction}, has recently attracted considerable attention in various research communities. Besides academic interest (see~\cite{Mohammadsurveylinkprediction, NowellCIKM2003} for a survey of different methods), link prediction has many important commercial applications, e.g., recommending friends in an online social network such as Facebook and suggesting potential hires in a professional network such as LinkedIn.

\begin{figure}
\centering
\includegraphics[width=\columnwidth]{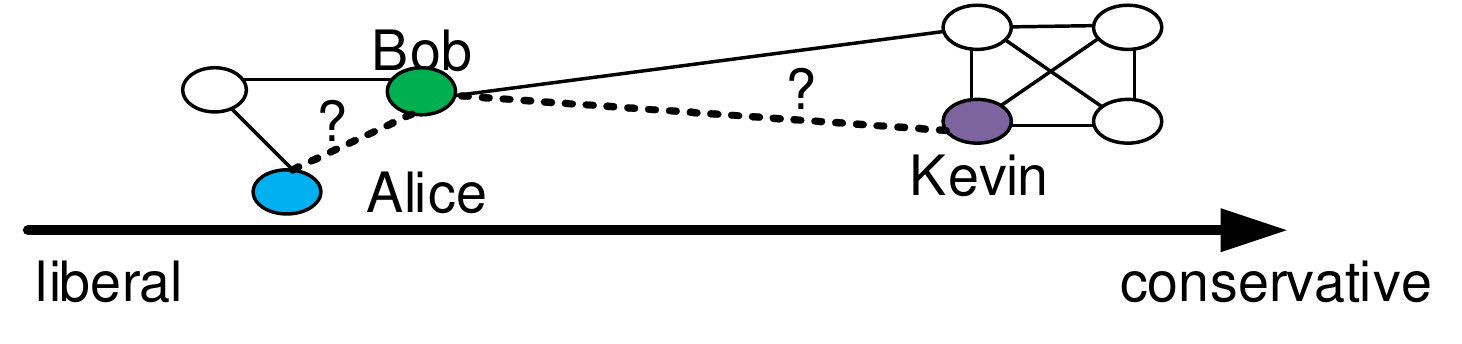}
\caption{\small{An example of observed interactions among Alice, Bob and Kevin, and their positions in a simplified one-dimension latent space. Alice is very liberal, Bob is biased towards being liberal, and Kevin is very conservative. However, all their profile information as well as the dimension label are unobservable.}} \label{fig:intuitionlatentspace}
\vspace{-0.4cm}
\end{figure}

In this work we focus on the temporal link prediction problem: Given a sequence of graph snapshots $G_1$, $\cdots$, $G_t$ from time 1 to $t$, how do we predict links in future time $t+1$? To perform link prediction in a network, one needs to construct models for link probabilities between pairs of nodes. Recent research interests in link prediction have focused on latent space modeling of networks. That is, given the observed interactions between nodes, the goal is to infer the position of each node in some latent space, so that the probability of a link between two nodes depends on their positions in that space. Latent space modeling allows us to naturally incorporate the well-known homophily effect~\cite{homophily} (birds of a feather flock together). Namely, each dimension of the latent space characterizes an unobservable homogeneous attribute, and shared attributes tend to create a link in a network. We illustrate this concept with an example shown in Figure~\ref{fig:intuitionlatentspace}.


\begin{example}
Assume that we have observed some interactions during a political campaign. An example of a latent space is shown in Figure~\ref{fig:intuitionlatentspace}, where most observed links were formed between users with similar political attitudes.
Based on the learned latent space, we could then predict that Bob is more likely to interact with Alice than Kevin in a
political campaign.
\end{example}

Various approaches, including Bayesian inference~\cite{Hoff2002,YinHX13}, multidimensional scaling (MDS)~\cite{Sarkarlatentspace2005}, matrix factorization~\cite{Menon:2011:LPV:2034117.2034146,QiICDE2013,GaoCIKM2011TLP,YeWWW2013,WWW13,ErdosTKDD2014}, and
 mixed-membership model~\cite{Airoldi:2008:MMS:1390681.1442798}, have been proposed to infer the static latent space
 representation of the observed network. Nevertheless, most of these studies were focusing on static graphs,
 where the latent positions of the nodes are fixed. In social networks, ``There is nothing permanent except change"; the
 network itself is dynamic and changing with time~\cite{VLDBGupta2014}~\cite{TantipathananandhKDD2007}. In addition, the latent position of each node might also be evolving over time~\cite{Sarkarlatentspace2005,Sun2007,DunlavyTKDDTLP,ZhuGCL14,Zhang:2014ICD}. A naive approach to extend static latent space modeling for dynamic networks is to model each node as a single latent representation and then update its position whenever the network evolves. Unfortunately, this approach tends to overfit on the current time step, leading to abrupt transitions and poor incorporation of historical information. Therefore, we are interested in inferring temporal latent positions for all the nodes in a dynamic network. The underlying problem can be stated as follows: \emph{Given a sequence of graph snapshots $G_1$, $\cdots$, $G_t$, how can we infer the temporal latent space so that at each time point links are more likely to be formed between nodes that are closer in the temporal latent space?} With the inferred temporal latent space, links can be accurately predicted at future times.

\begin{figure}
\centering
\includegraphics[width=\columnwidth]{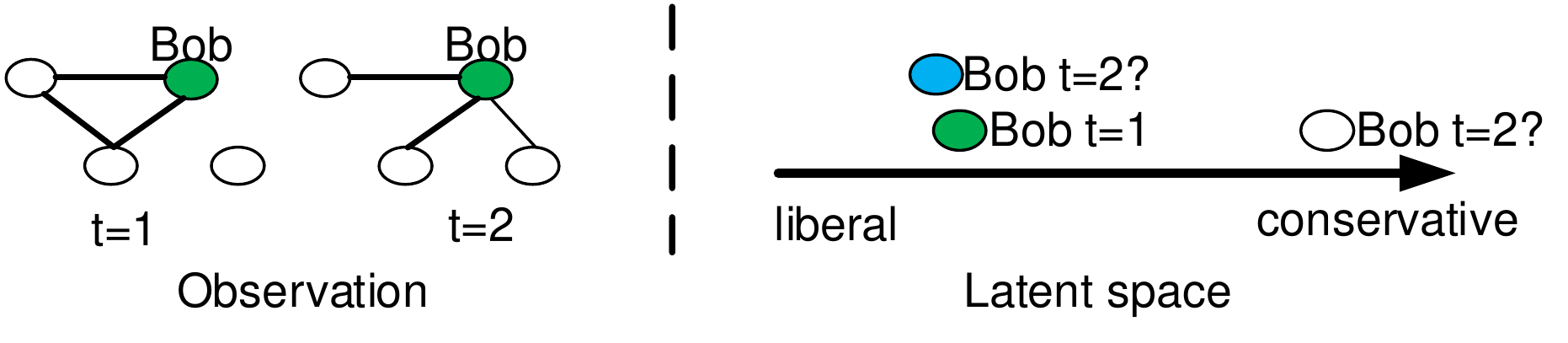}
\caption{\small{An example of temporal latent positions of user Bob. With the temporal smoothness assumption, Bob is more likely to be in the blue position than the white position in time $t=2$.}} \label{fig:temporallatentspace}
\vspace{-0.5cm}
\end{figure}

Unfortunately, we still have a limited understanding of how the latent space is evolving for temporal social networks.
In this work we propose a \emph{Temporal Latent Space Model for Dynamic Link Prediction}. Our model is based on the intuition
that nodes can move smoothly in the latent space over time, and that large moves are less likely~\cite{Sarkarlatentspace2005,Zhang:2014ICD}. As illustrated in Figure~\ref{fig:temporallatentspace}, in time $t=1$, Bob is biased towards liberal; it is unlikely that Bob will suddenly move to very conservative in time $t=2$. In addition to temporal smoothness, our model imposes a constraint on the dimensionality of the latent space and assumes that the dimension of latent space is much smaller than the number of nodes. With the dimensionality constraint, the online link prediction using our model is efficient in both computational time and storage cost (see related work for more details). In addition, varying the dimension of latent space offers an opportunity to fine-tune the compromise between computational cost and solution quality. The higher dimension leads to a more accurate latent space representation of each node, but also yields higher online computational cost.

%

One of the most widely used methods for inferring low-rank latent space in networks is via matrix factorization with the ``block-coordinate gradient descent (BCGD)" algorithm~\cite{Berry2007,TsengY09}, which has been successfully applied in community detection for
static networks~\cite{YangWSDM2013}. However, the introduced additional temporal smoothness term in the objective
function creates new challenges for the BCGD approach. Moreover, while there is a lot of work on
studying convergence of BCGD, it is unknown whether the new objective function satisfies these
criteria. We prove that the global BCGD algorithm proposed in this work has a quadratic convergence rate.

The global BCGD algorithm is computationally expensive and not applicable for truly large-scale problems.
Therefore, for the temporal link prediction problem with latent space models, a significant gap remains between theoretical
capabilities and practical applications. Towards this end, in this work we further introduce two new variants of
BCGD: a local BCGD algorithm and an incremental BCGD algorithm. In a local BCGD algorithm, instead of using all
historical graph snapshots to jointly infer temporal latent space at all timestamps, it sequentially infers the temporal latent space at each time position with a single graph snapshot and previous temporal latent space---thus significantly reducing computational cost. In addition, as we stated earlier, the latent positions of nodes are changing smoothly over time, not suddenly. To make use of this temporal smoothness, we develop an incremental BCGD algorithm to adaptively infer changes in latent positions based only on the changes in interactions. As shown in Figure~\ref{fig:temporallatentspace}, when Bob receives new interactions, his latent position in time $t=2$ requires an update which, however, can be performed efficiently leveraging his previous latent position in time $t=1$.


In summary, the contributions of this work are:
\begin{enumerate}
 \item We propose a temporal latent space model for dynamic networks to model the temporal link probability of node pairs, which can be further used to accurately recover or predict the formation of links.
 \item We address algorithmic issues in learning the temporal latent space representation of nodes by developing the standard global BCGD algorithm. We also provide a set of theoretical results for the standard global BCGD algorithm.
 \item We develop two fast BCGD algorithms: a local BCGD algorithm and an incremental BCGD algorithm. Furthermore, we illustrate that the proposed incremental BCGD algorithm only requires a conditional update on a small set of affected nodes, and is therefore significantly faster than all existing approaches.
 \item We conduct experiments over large real-life graphs. The experimental results show that the proposed approaches---global BCGD, local BCGD, and incremental BCGD---achieve good predictive quality in both sparse and dense networks with average AUC scores of 0.81, 0.79 and 0.78 respectively. In addition, it takes only less than half an hour for the incremental BCGD algorithm to learn the temporal latent spaces of massive networks with millions of nodes and links.
\end{enumerate}

The remainder of the paper is organized as follows: We first introduce our problem formulation in Section~\ref{sec:problem}, and then present the proposed global BCGD algorithm and two fast BCGD algorithms in Section~\ref{sec:bcgd} and Section~\ref{sec:fastbcgd} respectively. In Section~\ref{sec:expt}, we conduct an
experimental study to evaluate the effectiveness and efficiency of our
approach. Related work is discussed in Section~\ref{sec:related}. We
conclude this work and present future work in Section~\ref{sec:con}.

\section{Problem Formulation}\label{sec:problem}
A graph $G$ is denoted as ($V$, $E$), where $V$ is the set of nodes and $E\subseteq V\times V$
is the set of (directed or undirected) interactions. In this work we first focus on undirected graphs, in which case its matrix representation is symmetric. We use $u$ and $v$ to denote individual nodes, and $t$ and $\tau$ to denote timestamps. Let $G_{\tau}$ $=$ ($V_{\tau}$, $E_{\tau}$) be a time-dependent network snapshot recorded at time $\tau$, where $V_{\tau}$ is the set of nodes and $E_{\tau}\subseteq V_{\tau}\times V_{\tau}$ is the set of interactions. In addition, denote by
$\Delta V_{\tau}$ and $\Delta E_{\tau}$ the sets of vertices and interactions to be introduced (or removed) at time $\tau$, and let $\Delta G_{\tau}$ = ($\Delta V_{\tau}$, $\Delta E_{\tau}$) denote the change in the whole network.

\vspace{0.1cm}
\noindent \textbf{Dynamic social network.} A dynamic network $G$ is a sequence of
network snapshots within a time interval and evolving over time: $G$ $=$ ($G_1$, \dots, $G_t$).

\vspace{0.1cm}
\noindent \textbf{Temporal latent space and its model.} Let $Z_{\tau}$ be the low-rank $k$-dimension temporal latent space representation for node set $V_{\tau}$ at time $\tau$. For each individual $u$ at time $\tau$, we use a row vector $Z_{\tau}(u)$ to denote its temporal latent space representation and a scalar $Z(u, c)$ to denote its position in $c^{th}$ dimension. Inspired by the dynamic community model~\cite{TantipathananandhKDD2007,Tantipathananandh:2011:FCD:2117684.2118320}, we impose the following assumptions on the temporal latent space:
\begin{enumerate}
\item \textbf{Temporal smoothness}: Individuals change their latent positions gradually over time.
\item \textbf{Network embedding}: Two nodes that are close to each other in the interaction
network, in terms of the network distance, are also close to each other in the temporal latent space in terms of latent space distance.
\item \textbf{Latent Homophily}: Two members who are close to each other in latent space interact with one another more frequently than two faraway members.
\end{enumerate}
Within our model, we assume that the probability of a link between two nodes depends only on their latent positions. In principle, given a dynamic graph $G$ $=$ ($G_1$, \dots, $G_t$), we need $Z_{t+1}$ to predict future graph $G_{t+1}$. However, since $Z_{t+1}$ is not available, in our model we assume it can be approximated by $Z_1, \cdots, Z_{t}$. Therefore, given a dynamic graph $G$ $=$ ($G_1$, \dots, $G_t$), we infer $Z_1$, $\cdots$, $Z_{t}$ based on $G_1$, $\cdots$, $G_{t}$, use $Z_1$, $\cdots$, $Z_t$ to approximate $Z_{t+1}$, and finally predict $G_{t+1}$.

With the above definitions, we focus on the following problem:
\begin{problem}\label{problem:dcm}
\emph{(\textbf{Temporal Latent Space Inference.})} Giv\-en a dynamic social network $G$ = ($G_1$, $G_2$, \dots, $G_t$), we aim to find a $k$-dimension latent space representation at each timestamp $Z_{\tau}$ that minimizes the quadratic loss with temporal regularization:
 \begin{equation}\label{equ:objective}
 \small{
  \begin{aligned}
&\argmin_{Z_1, \cdots, Z_t}\sum_{\tau=1}^t\|G_{\tau}-Z_{\tau}Z_{\tau}^T\|_F^2+\lambda\sum_{\tau=1}^t\sum_{u}(1-Z_{\tau}(u)Z_{\tau-1}(u)^T)\\
&\mbox{subject to: $\forall$ $u$, $\tau$, }Z_{\tau}\geq 0, Z_{\tau}(u)Z_{\tau}(u)^T=1
 \end{aligned}
 }
 \end{equation}
where $\lambda$ is a regularization parameter, and the term $(1-Z_{\tau}(u)Z_{\tau+1}(u)^T)$ penalizes node $u$ for suddenly changing its latent position. Note that when computing the quadratic loss $\|G_{\tau}-Z_{\tau}Z_{\tau}^T\|_F^2$, we ignore all of the diagonal entries.
 \end{problem}

In the above model, the latent representation of each node corresponds to a point on the surface of a unit hypersphere. Note that this is different from mixed membership stochastic block models~\cite{Airoldi:2008:MMS:1390681.1442798} where nodes are mapped onto simplex. In practice, we find that sphere modeling gives us a clearer boundary between linked pairs and non-linked pairs when we project all pairs of nodes into the latent space. In addition, we impose the constraints $Z_{\tau}\geq 0$, not only because the non-negativity establishes the duality between our modeling and non-negative matrix factorization, but also because it gives latent space an intuitive parts-based interpretation, as suggested by Lee and Seung~\cite{lee99}. In the facial image example~\cite{lee99}, with non-negative constraints, each dimension of latent space corresponds to a part of the face, such as nose, eye and ear; while without non-negative constraints, each dimension of latent space corresponds to a blur representation of the entire face, of which it is very difficult to interpret. Similarly, in the social network domain, with the non-negative constraints, each dimension of latent space corresponds to a part of users' attributes such as current city, hometown, personality and so on. It is more intuitive to represent each user as an additive mixture of attributes (current city: A, hometown: B, personality: openness to experience), rather than represent each user as a combination of different representatives.

\vspace{0.1cm}
\noindent\textbf{Link prediction.}
Given that we have inferred $Z_1, \ldots, Z_t$ by optimizing Eq.~\ref{equ:objective}, our goal is to predict the adjacency matrix
$G_{t+1}$ at the next timestamp $t+1$. The most natural estimator is the conditional expectation: $Y_{t+1} = E[G_{t+1}\mid Z_1,\cdots, Z_t]$.
By assuming that the temporal dynamics of latent positions is Markovian and satisfies {\small
$E[Z_{t+1}\mid Z_1, \cdots, Z_t]=Z_t$ and $\mathrm{cov}[Z^T_{t+1}\mid Z_1,\cdots,Z_t] = D_t$} (a diagonal matrix), as
well as {\small $Z_{t+1}Z_{t+1}^T$} an unbiased estimate of {\small $G_{t+1}$}, we obtain
\begin{equation*}
\small{
\begin{aligned}
Y_{t+1}&=E[G_{t+1}\mid Z_1,\cdots, Z_t]\\
&=E[E[G_{t+1}\mid Z_1, \cdots Z_t, Z_{t+1}]\mid Z_1,\cdots, Z_t]\\
&=E[Z_{t+1}Z_{t+1}^T\mid Z_1, \cdots, Z_t]\\
&=\mathrm{cov}[Z^T_{t+1}\mid Z_1\cdots Z_t]\\
&+E[Z_{t+1}\mid Z_1\cdots Z_t]E[Z_{t+1}\mid Z_1\cdots Z_t]^T\\
&=D_t + Z_tZ_t^T.
\end{aligned}}
\end{equation*}
However, as we ignore the diagonal entries of the adjacency matrix (the graph does not contain self-loops) and because
the conditional covariance matrix $D_t$ is assumed to be diagonal, we will use the off-diagonal of {\small $Z_{t}Z_{t}^T$} to
predict $G_{t+1}$.


\vspace{0.1cm}
\noindent\textbf{Generalization.} Although in this work, we use {\small $Z_{t}Z_{t}^T$} to
predict $G_{t+1}$, without losing generality, the predicted graph at $t+1$ can be formulated as: $Y_{t+1}=\Phi (f(Z_1, \cdots, Z_t))$, where $\Phi$ is the link function and $f$ is the temporal function. Learning or selecting the best link function $\Phi$ and temporal function $f$ is beyond the scope of this work. For example, we could apply nonparametric approaches~\cite{SarkarCJICML12} to automatically learn the link function $\Phi$. Additionally, though this work focuses on undirected graphs, our model can be generalized to directed and weighted graphs. Specifically, this can be done by introducing another matrix $B$, which represents the weighted mapping from one dimension to another dimension in the latent space. With the matrix $B$, the term $G_{\tau}-Z_{\tau}Z_{\tau}^T$ in Eq.~\ref{equ:objective} is now generalized to $G_{\tau}-Z_{\tau}BZ_{\tau}^T$. If $B$ is symmetric, the underlying graph is undirected; otherwise, it is directed.

\vspace{0.1cm}
\noindent\textbf{Problem complexity.} Vavasis recently proved that non-negative matrix factorization (NMF) is NP-hard~\cite{Vavasis2009}. In addition, with the separability condition, that is, for each dimension $c$ in the latent space $Z$, there is a node $u$ such that $Z(u, c)\geq p$ and $Z(u, c^{\prime})=0$ for $c^{\prime}\neq c$, there exists a polynomial time exact algorithm to solve the non-negative matrix factorization problem~\cite{AroraSTOC2012}. Unfortunately, in our modeling, there is no guarantee that the latent space $Z$ satisfies the separability condition. In addition, even if there exists a latent space $Z$ that satisfies the separability condition, Recht et al.~\cite{RechtRTBNIPS12} pointed out that an exact algorithm is still too computationally expensive for large-scale data. Therefore, in the following we focus on the block coordinate gradient descent (BCGD) approach to obtain an approximate solution to Problem~\ref{problem:dcm}.

\section{A Standard BCGD Algorithm}\label{sec:bcgd}
In this section we present the details of the standard BCGD algorithm that provide a local optimal solution to Problem~\ref{problem:dcm}.

With a partial observed graph structure, the objective function in Eq.~\ref{equ:objective} can be decomposed into a linked part and a non-linked part. That is,
\begin{equation}\label{equ:dual1}
\small{
 \begin{aligned}
   &\argmin_{Z_1, \cdots,  Z_t}\sum_{\tau=1}^t\sum_{u,v\in E_{\tau}}(G_{\tau}(u,v)-Z_{\tau}(u)Z_{\tau}(v)^T)^2\\
   &+\sum_{\tau=1}^t\sum_{u,v\not\in E_{\tau}}(Z_{\tau}(u)Z_{\tau}(v)^T)^2\\
   &+\lambda\sum_{\tau=1}^t\sum_{u}(1-Z_{\tau}(u)Z_{\tau-1}(u)^T)\\
   &\mbox{subject to: $\forall$ $u$, $\tau$, }Z_{\tau}\geq 0, Z_{\tau}(u)Z_{\tau}(u)^T=1
 \end{aligned}
 }
\end{equation}

Unfortunately, the decomposed objective function in Eq.~\ref{equ:dual1} for Problem~\ref{problem:dcm} is still a fourth-order polynomial and non-convex. Therefore, we then adopt a block coordinate descent approach to solve Eq.~\ref{equ:dual1}. We update $Z_\tau(u)$ for each node $u$ at time $\tau$ by fixing both latent positions $Z_{\tau}(v)$ of all other nodes $v$ at time $\tau$ as well as all the temporal latent positions other than at time $\tau$. Proceeding in this way, each step of block coordinate descent is solving a convex problem.

For each node $u$ at time $\tau$, we focus on optimizing the following problem:
\begin{equation}\label{equ:subopt}
\small{
 \begin{aligned}
&\argmin_{Z_{\tau}(u)\geq 0} J(Z_{\tau}(u)), \mbox{ where } J(Z_{\tau}(u))=\\
&\sum_{v\in N(u)}(G_{\tau}(u,v)-Z_{\tau}(u)Z_{\tau}(v)^T)^2+\sum_{v\not\in N(u)}(Z_{\tau}(u)Z_{\tau}(v)^T)^2\\
&+\lambda(1-Z_{\tau+1}(u)Z_{\tau}(u)^T)+\lambda(1-Z_{\tau}(u)Z_{\tau-1}(u)^T)\}\\
 \end{aligned}
 }
\end{equation}
%
%

In the following, we use the projected gradient descent algorithm to find the approximation solution with a non-negativity constraint.
With the gradient descent optimization algorithm, for each node $u$ at timepoint $\tau$, we could iteratively update $Z_{\tau}$(u) in each iteration $r+1$ with the following rule:
\begin{equation*}
\small{
Z_{\tau}^{(r+1)}(u)=Z_{\tau}^{(r)}(u)-\eta \nabla_{Z_{\tau}(u)}J(Z_{\tau}^{(r)}(u))
}
\end{equation*}
where $\eta$ is the step size.

\vspace{0.1cm}
\noindent \textbf{Step size.} Nesterov's gradient method~\cite{opac-b1104789,journals/tsp/GuanTLY12,beck2009fast} iteratively updates the step size using the Lipschitz constant. According to the following lemma, we show that our step size can also be iteratively updated using the Lipschitz constant.

\begin{lemma}\label{lemma:lipschitz}
The gradient of the objective function $J(Z_{\tau}(u))$ is Lipschitz continuous, and the Lipschitz constant $L$ is equal to $2\sqrt{n^2-2n+k}$, where $n$ is the number of nodes in a graph, and $k$ is the number of dimensions. 
\end{lemma}
\noindent \textbf{Proof}: A detailed proof is in Supplement~\ref{app:1}. $\square$

With the Lipschitz constant and $\nabla_{Z_{\tau}(u)}J(Z_{\tau}^{(r)}(u))$ shown in Eq.~\ref{equ:particial} in Supplement~\ref{app:1}, each node $u$ at timepoint $\tau$ can be computed via the update rule stated in the following Lemma.

\begin{lemma}\label{lemma:updaterule}
The latent position of each node $u$ at timepoint $\tau$ can be iteratively computed with the following update rule:
\begin{equation}\label{equ:updateZ}
\small{
 \begin{aligned}
&Z_{\tau}^{(r+1)}(u)=max((1+2\alpha)Z_{\tau}^{(r)}(u)+\alpha\lambda (Z_{\tau-1}^{(r)}(u)+Z_{\tau+1}^{(r)}(u))\\
&+2\alpha\sum_{v\in N(u)}G_{\tau}(u,v)Z_{\tau}^{(r)}(v)-2\alpha Z_{\tau}^{(r)}(u){Z_{\tau}^{(r)}}^TZ_{\tau}^{(r)},0)
\end{aligned}
}
\end{equation}
where $\alpha=$ $\frac{a_{r+1}+a_r-1}{a_{r+1}L}$, $N(u)$ denotes the set of $u$'s neighbors, $L$ is the Lipschitz constant, $d(u)$ is the degree of node $u$, and
\begin{equation}\label{eq:ar}
a_r=\begin{cases} 1 & \text{if $r=0$,}\\
\frac{1+\sqrt{4a_{r-1}^2+1}}{2}&\text{if $r>0$.}
\end{cases}
\end{equation}
\end{lemma}

\noindent \textbf{Proof}: A detailed proof is in Supplement~\ref{app:2}. $\square$

Note that when updating $Z_{\tau}(u)$ according to Lemma~\ref{lemma:updaterule}, it requires the latent positions of all the neighbors of $u$. For nodes with extra-large degrees, the above update will become time-consuming. One interesting direction in the future is to apply stochastic gradient descent update rules to nodes with large degrees. That is, instead of using the latent positions of all of $u$'s neighbors, each iteration randomly selects one of $u$'s neighbors $v$, and leverages $v$'s latent positions to update the latent position of $u$.

We summarize our global block coordinate descent approach to solve Eq.~\ref{equ:dual1} in Algorithm~\ref{alg:bcgd1}. In the following, we provide additional theoretical properties of the proposed BCGD algorithm, including convergence analysis and complexity analysis.
\begin{algorithm}[!t]
\caption{The global BCGD algorithm for jointly inferring temporal latent space}\label{alg:bcgd1}
\small{
\begin{tabbing}
Input: Graphs $\{G_1$,$\cdots$, $G_t\}$ and latent space\\
\hspace{1.05cm}dimension $k$\\
Output: latent space $\{Z_1$,$\cdots$, $Z_t\}$ and prediction $Y_{t+1}$\\
1: Nonnegative initial guess for $\{Z_1$,$\cdots$, $Z_t\}$\\
2: \textbf{Repeat}\\
3: \hspace{0.5cm}\textbf{for} each time $\tau$ from 1 to $t$\\
4: \hspace{1.5cm}\textbf{for} each $u$ in graph\\
5: \hspace{2.0cm}update $Z_{\tau}(u)$ by Eq.~\ref{equ:updateZ}\\
6: \hspace{2.0cm}normalize $Z_{\tau}(u)$\\
7: \textbf{until} $\{Z_1$,$\cdots$, $Z_t\}$ converges.\\
8: \textbf{return} $Y_{t+1}=\Phi ([Z_1,\cdots, Z_t])$ and $\{Z_1$,$\cdots$, $Z_t\}$
\end{tabbing}
}
\vspace{-0.2cm}
\end{algorithm}
\subsection{Theoretical Analysis}
%
%
%
\noindent\textbf{Convergence rate.} Since our algorithm uses Nesterov's
gradient method to determine the step size (see Lemma~\ref{lemma:updaterule}), we can conclude that our algorithm achieves the convergence rate $O(\frac{1}{r^2})$ as stated in the following theorem. 

\begin{theorem}\label{theory:errorrate}
Given the sequence generated by Algorithm~\ref{alg:bcgd1}, for each node $u$ at timepoint $\tau$, we have $J(Z_{\tau}^{(r)}(u))-J(Z_{\tau}^*(u))$ $\leq\frac{2L\|Z_{\tau}^I(u)-Z_{\tau}^*(u)\|_F^2}{(r+2)^2}$, where $Z_{\tau}^*(u)$ is the optimum solution for $Z_{\tau}(u)$ with respect to the subproblem in Eq.~\ref{equ:subopt}, $Z_{\tau}(u)^I$ is the initialization for $Z_{\tau}(u)$, $L$ is the Lipschitz constant in Lemma~\ref{lemma:lipschitz}, and $r$ is the iteration number (not the total number of iterations).
\end{theorem}
\noindent \textbf{Proof}: The proof is shown in Supplement~\ref{app:3}. $\square$

\vspace{0.1cm}
\noindent \textbf{Local error bound.} The solution returned by Algorithm~\ref{alg:bcgd1} is a local optimum of the objective in Eq.~\ref{equ:dual1}. Unfortunately, it is not trivial to assess how this locally optimal solution compares to the global optimum of the objective in Eq.~\ref{equ:dual1}. This is because the objective function in Eq.~\ref{equ:dual1} is non-convex, and thus arriving at the global optimum via local iterations is not guaranteed. Only if the input matrix $G$ is separable or near-separable~\cite{AroraSTOC2012,RechtRTBNIPS12}, we are able to achieve the global optimum or a global error bound.

\begin{table}[!t]
  \centering
   \caption{\small{Time complexity analysis for Algorithm~\ref{alg:bcgd1}, $n$ is the number of nodes, $m_{\tau}$ is the number of edges in graph $G_{\tau}$, $d(u)$ is the degree of node $u$, $k$ is the number of dimensions, and $T$ is the number of timestamps.}}\label{tab:complexity}
\begin{tabular}{|c|c|c|}
  \hline
   & Sparse & Dense \\
   \hline
   Initialize&$O(\sum_{\tau}(n+m_{\tau})k)$  &O$(n^2Tk)$  \\
   \hline
   Update $Z_{\tau}(u)$& $O(d(u)k)$ & $O(nk)$ \\
   \hline
   Convergence&$O(\sum_{\tau}(n+m_{\tau}k)$  &$O(n^2Tk)$  \\
  \hline
\end{tabular}
\vspace{-0.3cm}
\end{table}

\vspace{0.1cm}
\noindent \textbf{Computational complexity.} Table~\ref{tab:complexity} summarizes the cost of each operation in an iteration for both sparse and dense graphs. Fortunately, Gibson et al.~\cite{Gibson2005} concluded that real-world networks are generally globally very sparse, where most of nodes have a relatively small number of neighbors. Therefore, for such real-world sparse networks, the total cost of a single iteration of BCGD should be $O(\sum_{\tau}(n+m_{\tau})k)$, which is \emph{linear in the number of edges and nodes}. Assume that the total number of iterations is $R$, then the time complexity is bounded by $O(Rk\sum_{\tau}(n+m_{\tau}))$. Since in Algorithm~\ref{alg:bcgd1}, we need to store all the input matrices and the output temporal latent space matrices, the storage complexity is bounded by $O(\sum_{\tau}(nk+m_{\tau}))$.

\section{Fast BCGD Algorithms}\label{sec:fastbcgd}
\begin{figure}
  \centering
  \subfigure[BCGD]{\includegraphics[width=0.49\columnwidth]{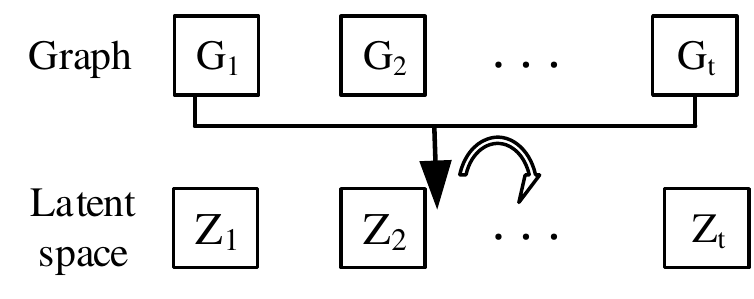}}
  \subfigure[Local BCGD]{\includegraphics[width=0.49\columnwidth]{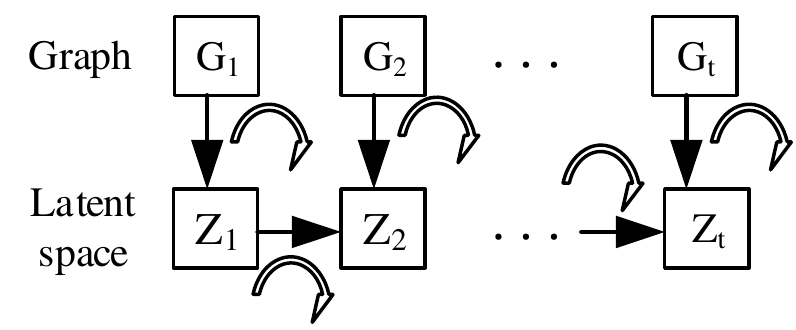}}
  \caption{\small{BCGD versus local BCGD.}}\label{fig:BCGD}
\end{figure}
Unfortunately, the standard BCGD is very expensive in both computation and storage since it requires all historical graph snapshots $\{G_1$,$\cdots$, $G_t\}$ to jointly and cyclicly update all temporal latent space $\{Z_1$, $\cdots$, $Z_t\}$. In this section we aim to further reduce the time complexity of the BCGD algorithm so that it scales with big data. In contrast to the joint inference in BCGD, we propose a sequential inference algorithm and an incremental inference algorithm, both of which utilize the locality and temporal information. The proposed two faster BCGD algorithms are as efficient as the fast independent inference (i.e., infer $Z_{\tau}$ from $G_{\tau}$ only) and as effective as the global joint inference (BCGD).

As illustrated in Figure~\ref{fig:BCGD}, we first introduce the local BCGD algorithm, which sequentially infers each temporal latent space $Z_{\tau}$ from a single graph snapshot $G_{\tau}$ and partial prior knowledge $Z_{\tau-1}$.

\subsection{Local BCGD Algorithm}
Specifically, at each timestamp $\tau$, we aim to optimize the following local objective function to compute $Z_{\tau}$:
\begin{equation}\label{equ:subopt2}
\small{
\begin{aligned}
&\argmin_{Z_{\tau}}\sum_{u, v\in E_{\tau}}(G_{\tau}(u,v)-Z_{\tau}(u)Z_{\tau}(v)^T)^2\\
&+\sum_{u, v\not\in E_{\tau}}(Z_{\tau}(u)Z_{\tau}(v)^T)^2\\
&+\lambda \sum_{u\in V_{\tau}}(1-Z_{\tau}(u)Z_{\tau-1}(u)^T)
\end{aligned}}
\end{equation}

Using the same BCGD approach, we iteratively update the latent position $Z_{\tau}(u)$ of each node $u$ by fixing the latent positions of all the other nodes. This leads to the following update rules for $Z_{\tau}(u)$ in the $(r+1)^{th}$ iteration:
\begin{equation}\label{equ:updateZ2}
\small{
\begin{aligned}
Z_{\tau}^{(r+1)}(u)=&\max((1+2\alpha)Z_{\tau}^{(r)}(u)+\alpha\lambda Z_{\tau-1}(u)+2\alpha\sum_{v\in N(u)}\\
&G_{\tau}(u,v)Z_{\tau}^{(r)}(v)-2\alpha Z_{\tau}^{(r)}(u){Z_{\tau}^{(r)}}^TZ_{\tau}^{(r)}, 0)
\end{aligned}}
\end{equation}
where $\alpha=$ $\frac{a_{r+1} + a_{r} - 1}{a_{r+1}L}$, $L$ is the Lipschitz constant, and $a_r$ is defined in Lemma~\ref{lemma:updaterule}.

\begin{algorithm}[!t]
\caption{The local BCGD algorithm for sequentially inferring temporal latent space}\label{alg:bcgd2}
\small{
\begin{tabbing}
Input: Graphs $\{G_1$,$\cdots$, $G_t\}$ and latent space\\
\hspace{1.05cm}dimension $k$\\
Output: $Y_{t+1}$ and latent space $\{Z_1$,$\cdots$, $Z_t\}$\\
1: Nonnegative initial guess for $Z_1$\\
2: \textbf{for} each $\tau$ from 1 to $t$\\
3: \hspace{0.5cm}Initial $Z_{\tau}$ based on $Z_{\tau-1}$\\
4: \hspace{0.5cm}\textbf{repeat}\\
5: \hspace{1.0cm}\textbf{for} each $u$ in graph $G_{\tau}$\\
6: \hspace{1.5cm}update $Z_{\tau}(u)$ by Eq.~\ref{equ:updateZ2} and normalize it\\
7: \hspace{0.5cm}\textbf{until} $Z_{\tau}$ converges.\\
8: \textbf{return} $Y_{t+1}=\Phi ([Z_1,\cdots, Z_t])$ and $\{Z_1$,$\cdots$, $Z_t\}$
\end{tabbing}
}
\vspace{-0.2cm}
\end{algorithm}
We summarize the proposed local BCGD algorithm in Algorithm~\ref{alg:bcgd2}. Note that in the global BCGD algorithm, for each iteration we jointly update $Z_1$, $Z_2$  until $Z_t$ and then iterate back to $Z_1$ in the next iteration. This cyclic update schema is very expensive; while in the local BCGD algorithm, as shown in Lines 4--7, we sequentially compute $Z_{\tau}$ by a single inner iteration. That is, we iteratively update $Z_{\tau}$ until $Z_{\tau}$ converges and then move to the computation of temporal latent space $Z_{\tau+1}$. This local sequential update schema greatly reduces the computational cost in practice, as we analytically show in the following.

\noindent\textbf{Complexity analysis.} The cost of each operation in Algorithm~\ref{alg:bcgd2} remains the same as that of Algorithm~\ref{alg:bcgd1}, as reported in Table~\ref{tab:complexity}. Thus, the total computational cost is $O(\sum_{\tau}(n+m_{\tau})R_{\tau}k)$. In practice, since we leverage prior knowledge $Z_{\tau-1}$ to locally update $Z_{\tau}$, it converges much faster than the global BCGD algorithm, and thus the local BCGD algorithm is more efficient than the global BCGD algorithm, as also verified in our experiments. In addition, in the in-memory algorithm where we put all the data into memory, the global BCGD algorithm requires at least $O(\sum_{\tau}(nk+m_{\tau}))$ memory cost; while the local BCGD only requires storage of a single graph snapshot and two latent space matrices representation. Therefore, the memory cost of the local BCGD algorithm is bounded by $O(nk+\max_{\tau}m_{\tau})$.

\subsection{Incremental BCGD Algorithm}
In this section we further study how to infer or maintain temporal latent space incrementally with graph changes (nodes and edges insertion and deletion). Instead of recomputing the temporal latent space $Z_{\tau}$ with the entire graph snapshot $G_{\tau}$ and the prior knowledge $Z_{\tau-1}$, our dynamic update strategy is to adjust $Z_{\tau}$ incrementally from $Z_{\tau-1}$ as the network structure changes. Specifically, we take advantage of what we already learned in previous snapshots, as well as the temporal transition for inferring the dynamics in the current timestamp.

\noindent\textbf{Overview of incremental BCGD.} The overview of this algorithm is as follows: First, we identify nodes whose local neighborhood has changed between two consecutive snapshots, including cases where an existing node adds or deletes links, or a node is added to or removed from the network. With the matrix representation, we model all the four updates towards $G_t$ in terms of row-wise updates on matrix representation. For example, a newly added node with degree $d$ is modeled as an update from a row with all-zero entries to a row with $d$ non-zero entries. For nodes whose local neighborhood has not changed, we assume that their initial temporal latent positions do not change either. For nodes whose neighborhood has changed, we initialize their new temporal latent positions based on their new neighbors' latent space positions. Next, we iteratively perform a conditioned latent position update for each affected node (i.e., a candidate set of nodes whose latent position might change) and an update for the affected node set until there are no more changes in the temporal latent position of each node. The entire process is outlined in Algorithm~\ref{alg:bcgd3}.

\begin{algorithm}[!t]
\caption{The incremental BCGD algorithm for inferring temporal latent space}\label{alg:bcgd3}
\small{
\begin{tabbing}
Input: Graphs $\{G_1$,$\cdots$, $G_t\}$ and latent space\\
\hspace{1.05cm}dimension $k$\\
Output: $Y_{t+1}$ and latent space $\{Z_1$,$\cdots$, $Z_t\}$\\
01: Nonnegative initial guess for $Z_1$\\
02: \textbf{for} each time stamp $\tau$ from 1 to $t$\\
03:\hspace{0.5cm}\textbf{for} each $u$ in graph $G_{\tau}$\\
04:\hspace{1.0cm}\textbf{if} $G_{\tau}(u)$ is not updated\\
05:\hspace{1.5cm}$Z_{\tau}(u)=Z_{\tau-1}(u)$\\
06:\hspace{1.0cm}\textbf{else}\\
07:\hspace{1.5cm}Initialize $Z_{\tau}(u)$ using Eq.~\ref{equ:initialZu}\\
08:\hspace{0.5cm}affected node set $S$=$\Delta V_{\tau}$\\
09:\hspace{0.5cm}\textbf{repeat}\\
10:\hspace{1.0cm}\textbf{for} each $u$ in $S$\\
11:\hspace{1.5cm}update $Z_{\tau}(u)$ by Eq.~\ref{equ:updateZ2} and normalize it\\
12:\hspace{1.0cm}update affected node set $S$ with Alg.~\ref{alg:nodeselection}\\
13:\hspace{0.5cm}\textbf{until} $Z_{\tau}$ converges or $S$ is empty.\\
14: \textbf{return} $Y_{t+1}=\Phi ([Z_1,\cdots, Z_t])$ and $\{Z_1$,$\cdots$, $Z_t\}$
\end{tabbing}
}
\vspace{-0.2cm}
\end{algorithm}
\noindent \textbf{Initialize updated nodes.} In our algorithm, the four update operations are handled simply by comparing whether $G_{\tau}(u)$ is identical to $G_{\tau-1}(u)$. For each updated node $u$, we initialize its latent position based on the probability of seeing its updated neighbors' latent positions. Specifically, for each node $u$ and the dimension $c$ of the latent space at time $\tau$, the position of $u$ in dimension $c$ is computed using the following equation:

\begin{equation}\label{equ:initialZu}
Z_{\tau}(u,c)=\frac{\sum_{v\in N(u)}G_{\tau}(u,v)Z_{\tau}(v,c)}{\sum_{q\in N(u)}G_{\tau}(u,q)}
\end{equation}
The initialization of latent position for an updated node $u$ follows the notion of ``latent homophily" introduced earlier: The latent position of the node $u$ is as close as possible to those of its network neighbors.

\vspace{0.1cm}
\noindent \textbf{Identifying affected nodes.} Our dynamic update strategy can be viewed as an extra conditional update by which only nodes affected accept a new latent position. Unfortunately, the set of affected nodes for which the latent positions need to be updated is not limited to only the set of recently changed nodes and/or their network neighbors. Therefore, how could we identify the set of affected nodes?

The overall idea of our affected nodes identification is outlined as follows. Initially, the set of affected nodes is identical to the set of updated nodes (Line 8 in Algorithm~\ref{alg:bcgd3}). Next, after performing each iteration of a conditional update with Eq.~\ref{equ:updateZ2} in Line 11 in Algorithm~\ref{alg:bcgd3}, some old affected nodes are no longer required to be updated since their latent positions have converged. On the other hand, the effects of the update could be further propagated from old affected nodes to their neighborhood. The details of our affected nodes update procedure are presented in Algorithm~\ref{alg:nodeselection}.



\begin{algorithm}
\caption{Updating affected nodes}\label{alg:nodeselection}
\small{
\begin{tabbing}
Input: $Z_{\tau-1}$, $Z_{\tau}$, $S^{\texttt{old}}$\\
Output: A set of affected nodes $S$\\
1: $S=S^{\texttt{old}}$\\
2: \textbf{for} each $u$ in $S^{old}$\\
3: \hspace{0.5cm}\textbf{if} $\forall c$, $|Z_{\tau}(u,c)$ $-$ $Z_{\tau-1}(u,c)|<\delta$\\
4: \hspace{1.0cm}$S=S\setminus \{u\}$\\
5: \hspace{0.5cm}\textbf{for} each $w\in N(u)$\\
6: \hspace{1.0cm}\textbf{if} $|Z_{\tau}(u)Z_{\tau}(w)^T-Z_{\tau-1}(u) Z_{\tau-1}(w)^T|\geq\zeta$\\
7: \hspace{1.5cm}$S=S\cup \{w\}$\\
8: \textbf{return} $S$
\end{tabbing}
}
\vspace{-0.2cm}
\end{algorithm}

\vspace{0.1cm}
\noindent \textbf{Complexity analysis.} The total computational cost of Algorithm~\ref{alg:bcgd3} is $O(k\sum_{\tau}\sum_{r}|S_r+N(S_r)|)$, where $r$ denotes the iteration number of the conditional update process outlined in Lines 9--13 in Alg.~\ref{alg:bcgd3}, $S_r$ denotes the set of affected nodes and $N(S_r)$ denotes the neighborhood of affected nodes. When $\tau$=1, $|S_r|$ is equal to the number of nodes $n$, and $|N(S_r)|$ is equal to the number of edges in the first graph snapshot $m_1$. For the in-memory storage cost, compared with Algorithm~\ref{alg:bcgd2}, Algorithm~\ref{alg:bcgd3} requires additional storage cost about $\Delta G_{\tau}$. Therefore, the in-memory storage cost of Algorithm~\ref{alg:bcgd3} is $O(nk+\max_{\tau}(m_{\tau}+\Delta m_{\tau}))$.

\section{Experiments}\label{sec:expt}
\subsection{Dataset and Evaluation}
We use five real temporal datasets in our experiments, which are obtained from the Koblenz Large Network Collection~\cite{KONECT}. The statistics of each data set are reported in Table~\ref{tab:data}. Note that in our experimental setting, each graph snapshot $G_{\tau}$ consists of all the nodes and the interactions that appear within the time interval $[\tau-1,\tau]$.

\begin{table}[!t]
  \centering
  \caption{{\small Statistics of data sets, where volume denotes the total number of interactions, and the number of edges denotes the number of distinct interactions.}}\label{tab:data}
  \vspace{-0.2cm}
  \small{
  \begin{tabular}{|p{0.5in}|l|l|l|p{0.4in}|}
    \hline
    Data & \# nodes&volume& \# edges &\# snapshots\\
    \hline
    Infection \cite{konect:sociopatterns}& 410  &17,298 &2,765& 8\\
    \hline
    Facebook \cite{viswanath09}&63,731 &817,035&183,412& 5 \\
    \hline
    HepPh \cite{konect:leskovec107}&28,093  & 4,596,803 &3,148,447&9  \\
    \hline
    DBLP \cite{konect:2015:dblp_coauthor}&1,314,050  &18,986,618  &10,724,828& 11 \\
\hline
    YouTube \cite{konect:mislove2}&3,223,589  &9,375,374  &9,375,374&  7\\
    \hline
  \end{tabular}
  }
  \vspace{-0.3cm}
\end{table}

\vspace{0.1cm}
\noindent\textbf{Evaluation metrics.} We evaluate the performance of the proposed approaches from three aspects: the effect of parameters, the efficiency of both offline latent space inference and online link prediction, and the accuracy of online link prediction. We use total running time and memory consumption to evaluate the efficiency of both offline latent space inference and online link prediction. We use prediction error to evaluate the inference accuracy. Given the training graph $G_1$, $\cdots$, $G_t$, prediction error is defined as $\frac{1}{t-1}\sum_{\tau=2}^t\|G_{\tau}-Z_{\tau-1}Z_{\tau-1}^T\|_F$. Therefore, a smaller prediction error indicates better inference accuracy.

For link prediction accuracy, we use Area Under Curves (both Receiver Operating Characteristic (ROC) and Precision-Recall (PR) curves), termed as $\texttt{AUC}_{\texttt{ROC}}$ and $\texttt{AUC}_{\texttt{PR}}$. The ROC curve is created by plotting the true positive rate vs. false positive rate; while the PR curve is created by plotting precision vs. recall at various threshold settings. Thus, AUC score evaluates the overall ranking yielded by the algorithms with a larger AUC indicating better link prediction performance. In the following we only report the experiment results based on $\texttt{AUC}_{\texttt{ROC}}$; the experiment results based on $\texttt{AUC}_{\texttt{PR}}$ are reported in the Supplement.

\vspace{0.1cm}
\noindent\textbf{Test pair generation.} Given the training graph $G_1$, $\cdots$, $G_t$, we perform link prediction for the test graph $G_{t+1}$. We provide two different setups to generate test pairs: (1) {\em all links}: In this setup, we focus on how well the methods predict links in the test graph, no matter whether those links are repeated links or new links. Toward this goal, we randomly generate 100,000 linked and an equal number of non-linked pairs from $G_{t+1}$ as test pairs. (2) {\em new links}: In this setup, we focus on how well the methods predict the emergence of new links and deletion of existing links. Again, we randomly generate 10,000 linked and an equal number of non-linked pairs from $\Delta G_{t+1}$.

\vspace{0.1cm}
\noindent\textbf{Comparable approaches.} We compare our approaches (standard BCGD $\texttt{BCGD}_G$, local BCGD $\texttt{BCGD}_L$ and incremental BCGD $\texttt{BCGD}_I$) with a set of state-of-the-art approaches, which are summarized in Table~\ref{tab:baseline} and Section~\ref{sec:related}. For the offline latent space inference time comparison, we compare our approaches with global optimization approach DMMSB \cite{Fu:2009:DMM:1553374.1553416}, local optimization approaches NMFR \cite{YangnIPS12}, Hott \cite{RechtRTBNIPS12}, BIGCLAM \cite{YangWSDM2013}, PTM \cite{YinHX13} and incremental optimization approach LabelRT \cite{Xiecorr}. We compare online link prediction times of our approaches to that of high-dimension latent space approaches BIGCLAM and LabelRT, and popular graph heuristic AA~\cite{Adamic01friendsand}. For the Hott, NMFR and DMMSB, the total running time for their approaches is the same as that of our BCGD approaches. This is because the online prediction time cost is proportional to the number of dimensions $k,$ and $k$ is set to 20 for all the low-dimension constraint latent space approaches including our BCGD approaches, Hott, NMFR, PTM and DMMSB.

\begin{table*}
  \centering
  \small{
  \caption{A summary of state-of-the-art approaches, where \textit{time} \textit{series} denotes using a sequence of graph snapshots as inputs, \textit{aggregated} denotes using a single aggregated static graph snapshot as an input, \textit{low} denotes low dimensional latent space and \textit{high} denotes without dimensionality constraint.}\label{tab:baseline}
  \vspace{-0.2cm}
  \begin{tabular}{|c|c|c|c|c|c|c|c|c|c|}
    \hline
    &\multicolumn{3}{c|}{Latent space dimension}&\multicolumn{2}{c|}{Input graph}&\multicolumn{2}{c|}{Scalability}&\multicolumn{2}{c|}{Accuracy}\\
    \hline
    &low&high&NA&aggregated&time-series&inference&prediction&all links&new links\\
    \hline
    PTM~\cite{YinHX13}, NMFR~\cite{YangnIPS12}&$\surd$&&&$\surd$&&&$\surd$&$\surd$&\\
    \hline
    Hott~\cite{RechtRTBNIPS12}&$\surd$&&&$\surd$&&$\surd$&$\surd$&$\surd$&\\
    \hline
    LabelRT~\cite{Xiecorr}&&$\surd$&&$\surd$&$\surd$&&&&$\surd$\\
    \hline
    BIGCLAM~\cite{YangWSDM2013}&&$\surd$&&$\surd$&&$\surd$&&&\\
    \hline
    AA~\cite{Adamic01friendsand}&&&$\surd$&$\surd$&&NA&&$\surd$&\\
    \hline
    $\texttt{BCGD}_G$, DMMSB~\cite{Fu:2009:DMM:1553374.1553416}&$\surd$&&&$\surd$&$\surd$&&$\surd$&$\surd$&$\surd$\\
    \hline
    $\texttt{BCGD}_L$, $\texttt{BCGD}_I$&$\surd$&&&&$\surd$&$\surd$&$\surd$&$\surd$&$\surd$\\
    \hline
  \end{tabular}
  }
  \vspace{-0.3cm}
\end{table*}

\vspace{0.1cm}
\noindent\textbf {Configurations.} The results are reported as an average over 10 runs. If the maximum number of iterations is required as an input; we set it to 100. All experiments are conducted on a single machine with 8G memory and i7 2.7 GHZ CPU. 

\subsection{Effect of Parameters}
\begin{table}[t]
\centering
\caption{\small{Effect of the smoothing parameter $\lambda$}}\label{tab:lambda}
\vspace{-0.2cm}
\small{
\begin{tabular}{|c|c|c|c|}
  \hline
  \multicolumn{4}{|c|}{Prediction error by $\texttt{BCGD}_G$}\\
  \hline
  $\lambda$&0&$[0.0001,1]$&10\\
  \hline
Infection&228&225$\pm$1.26&288\\
\hline
Facebook&3436&3312$\pm$16&3410\\
\hline
Hepph&2043&1697$\pm$7.9&1769\\
\hline
DBLP&80566&65878$\pm$429&65926\\
\hline
Youtube&195023&161239$\pm$726&162282\\
  \hline
 \hline
  \multicolumn{4}{|c|}{Prediction error by $\texttt{BCGD}_L$}\\
  \hline
  $\lambda$&0&$[0.0001,1]$&10\\
  \hline
  Infection&250&243$\pm$3.5&249\\
\hline
Facebook&3659&3485$\pm$36&3525\\
\hline
Hepph&2566&2106$\pm$51&2172\\
\hline
DBLP&81633&65957$\pm$18&66471\\
\hline
Youtube&198646&162287$\pm$357&163675\\
  \hline
\end{tabular}
}
\vspace{-0.3cm}
\end{table}

This set of experiments aims to determine the impact of parameters on the performance of the proposed approaches and/or the comparable approaches.

\vspace{0.1cm}
\noindent\textbf{The effect of temporal smootheness. }While fixing all the other parameters (e.g., $k$=20), we first study the effect of the smoothing parameter $\lambda$. The effect of parameter $\lambda$ might be related to the inference algorithms, thus we evaluate the performance of both global BCGD and local BCGD when varying $\lambda$. We do not plot the running time since we notice that the running time is insensitive to the parameter $\lambda$. We vary $\lambda$ from 0.0001 to 10 with logarithmic scale and report the prediction errors shown in Table~\ref{tab:lambda}. We also report the prediction errors for $\lambda$=0. Note that when $\lambda$ is 0,  the update rules stated in Eq.~\ref{equ:updateZ} and Eq.~\ref{equ:updateZ2} become identical. $\texttt{BCGD}_G$ differs from $\texttt{BCGD}_L$ since they have different ways to proceed with initialization and convergence, see Lines 2--7 in Algorithm~\ref{alg:bcgd1} and~\ref{alg:bcgd2}. Clearly, either absence of temporal smoothness ($\lambda$=0) or strong temporal smoothness ($\lambda$=10) leads to higher prediction errors. While $\lambda$ varies from 0.0001 to 1, there are no significant differences in terms of prediction errors. Therefore, in the following experiments, we simply fix $\lambda$ as 0.0001 for $\texttt{BCGD}_G$ and 0.01 for $\texttt{BCGD}_L$.

\begin{figure}[!t]
  \centering
  \includegraphics[width=\columnwidth]{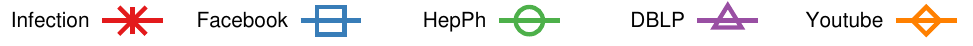}
  \subfigure[Inferring time (s)]{\includegraphics[width=0.492\columnwidth]{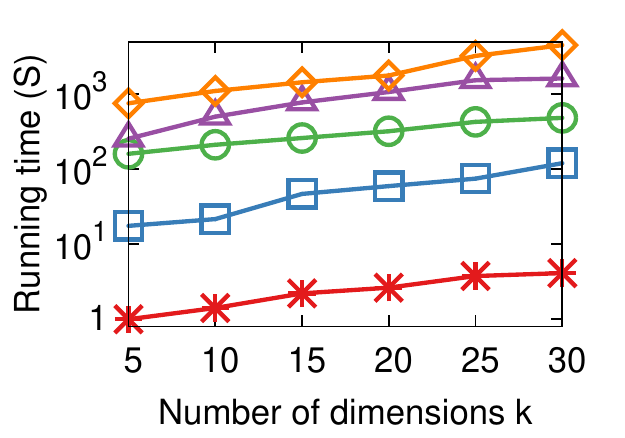}}
  \subfigure[Prediction errors]{\includegraphics[width=0.492\columnwidth]{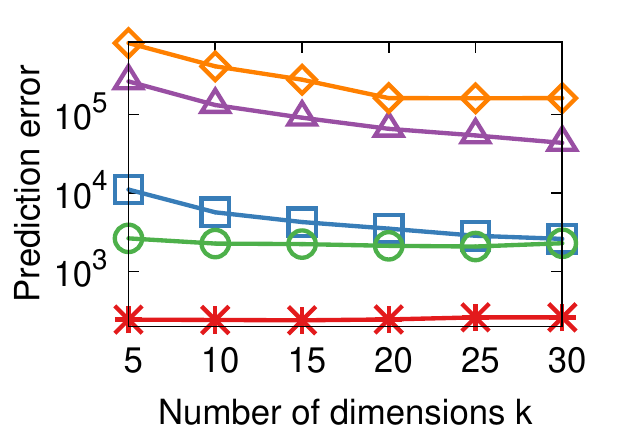}}
  \caption{\small{Effect of dimensionality $k$ to proposed approaches.}}\label{fig:K}
  \vspace{-0.3cm}
\end{figure}

\vspace{0.1cm}
\noindent\textbf{The effect of dimensionality. }
Next, we fix all the other parameters and analyze the effect of $k$. Since the effect of number of dimensions $k$ is lightly correlated with inference algorithms, here we choose the representative local BCGD algorithm. We vary $k$ from 5 to 30, and report both the running time of temporal latent space inference and prediction error in Figure~\ref{fig:K}. The overall trends indicate that the running time increases with number of dimensions $k$, while prediction error decreases with number of dimensions $k$. However, increasing number of dimensions does not necessarily lead to the decrease of prediction error. For example, for a Facebook dataset, when the number of dimensions $k$ is increased from 25 to 30, the prediction error is increased. In order to balance the efficiency and effectiveness, we opt to fix $k$=20 for the proposed approach in all of the following experiments. A nonparametric approach that automatically selects the best dimension $k$ will be an interesting future direction.

\begin{figure}[!t]
  \centering
  \includegraphics[width=\columnwidth]{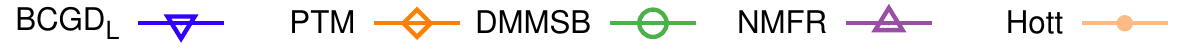}
  \subfigure[Inferring time (s)]{\label{fig:timeallk}\includegraphics[width=0.492\columnwidth]{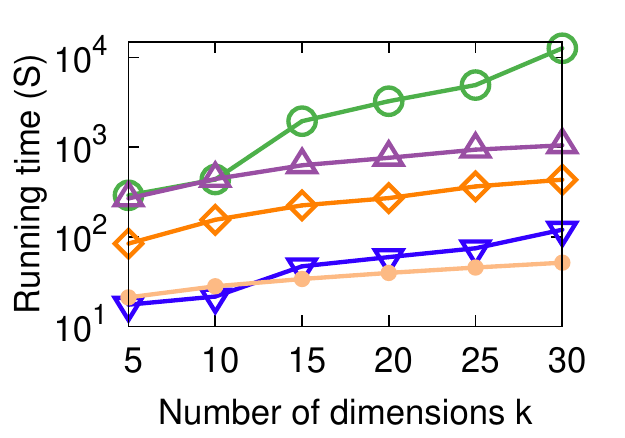}}
  \subfigure[Prediction accuracy]{\label{fig:rocallk}\includegraphics[width=0.492\columnwidth]{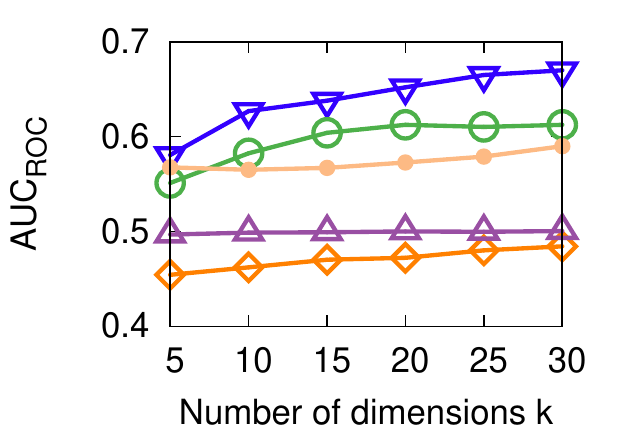}}
  \caption{\small{Effect of dimensionality $k$ to other approaches on Facebook dataset.}}\label{fig:K}
  \vspace{-0.3cm}
\end{figure}
In order to perform a fair comparison, we also examine the effect of $k$ to other comparable approaches including DMMSB, PTM, NMFR, and Hott. LabelRT and BIGCLAM are not included because their best dimension $k$ is automatically computed. Furthermore, since several baselines DMMSB, PTM, NMFR are not scalable and can not support running on the two large datasets DBLP and YouTube, we simply choose the relatively largest dataset, Facebook, on which all of the approaches are able to be tested. The comparison of running time and $\texttt{AUC}_{\texttt{ROC}}$ on new links is reported in Figure~\ref{fig:timeallk} and Figure~\ref{fig:rocallk} respectively. Clearly, the overall trends of other approaches are very similar to those of $\texttt{BCGD}_L$: The inference time is growing linearly or even quadratically with the dimensionality $k,$ while the prediction accuracy also improves with $k$. Therefore, we use the same value of $k$=20 for other comparable approaches as well.

\begin{figure}[!t]
  \centering
  \includegraphics[width=\columnwidth]{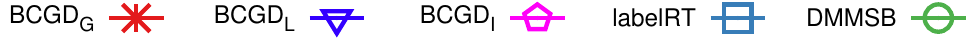}
  \subfigure[Inferring time (s)]{\label{fig:timetime}\includegraphics[width=0.492\columnwidth]{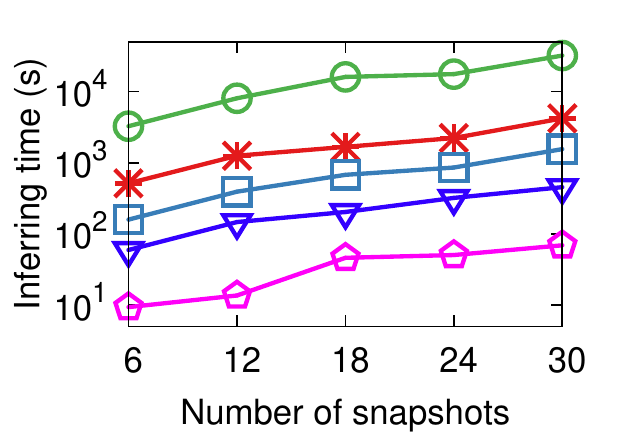}}
  \subfigure[Prediction accuracy]{\label{fig:roctime}\includegraphics[width=0.492\columnwidth]{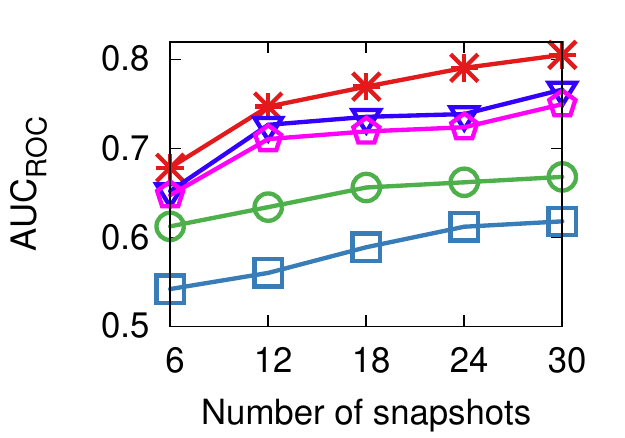}}
  \caption{\small{Effect of number of snapshots to proposed and other approaches on Facebook dataset.}}\label{fig:time}
  \vspace{-0.3cm}
\end{figure}

\vspace{0.1cm}
\noindent\textbf{The effect of time intervals. } In order to understand the proposed model more deeply, we also study how the value of time interval $\tau$ impacts prediction performance. Specifically, we choose the Facebook dataset, vary the value of time interval $\tau$, and create five different sequences of graph snapshots. As reported in Figure~\ref{fig:timetime} and Figure~\ref{fig:roctime}, both the inference time and prediction accuracy increase with number of snapshots. The inference time increases because a larger number of snapshots mean more frequent model updates per shorter time interval. Meanwhile, shorter time interval leads to less change of graphs, which indicates that it is more reliable to forecast links at time $t+1$ based on the current observation at time $t$. Therefore, the prediction accuracy also improves with number of snapshots. Again, the overall trends hold for both proposed and other approaches. Therefore, we simply fix the number of snapshots as reported in Table~\ref{tab:data}.

\begin{figure}[!t]
  \centering
  \subfigure[Inferring time (s)]{\includegraphics[width=0.492\columnwidth]{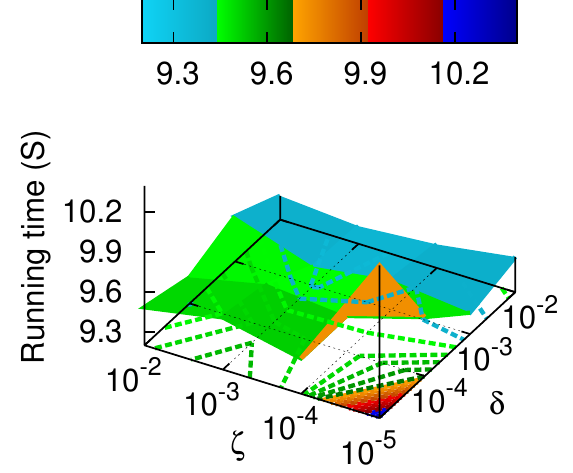}}
  \subfigure[Prediction errors]{\includegraphics[width=0.492\columnwidth]{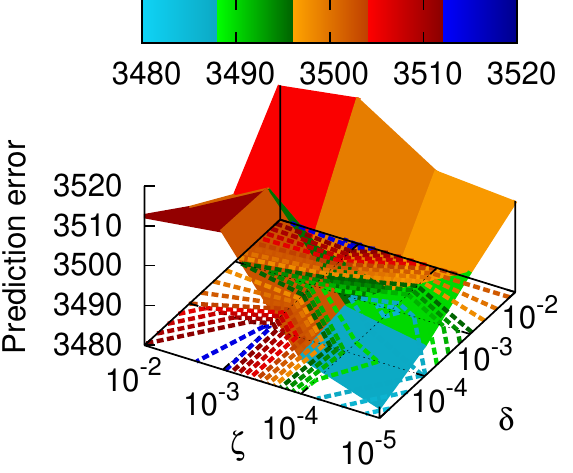}}
  \caption{\small{Effect of parameters $\zeta$ and $\delta$ on Facebook dataset (Best viewed in color).}}\label{fig:zeta}
  \vspace{-0.3cm}
\end{figure}

\vspace{0.1cm}
\noindent\textbf{The effect of lazy update. } Finally, we evaluate the effect of two parameters $\zeta$ and $\delta$ that are related to the incremental BCGD algorithm. Since parameters $\zeta$ and $\delta$ are correlated with each other, we vary both of them and report the total inference time and prediction error on the Facebook dataset in Figure~\ref{fig:zeta}. The results verify that when setting both $\zeta$ and $\delta$ to smaller values, the predictive accuracy is improved, while the total running time is also increased. However, Figure~\ref{fig:zeta}(a) shows that when we reduce $\zeta$ and $\delta$ from $10^{-2}$ to $10^{-4}$, the running time almost remains the same. The results verified that we can safely set $\zeta$ and $\delta$ to relatively small values without losing too much efficiency. We repeated this set of experiments on all the other datasets. Then, using curve fitting to capture how the best value of $\zeta$ and $\delta$ can be approximated by parameters such as $n$, $m$, $T$, and $k$, we noticed that a good value for $\zeta$ and $\delta$ can be achieved around $\sqrt{-\log(1-1/n)}$ and $2\zeta/k$. Therefore, in the following experiments, without specification, we set $\zeta$ as $\sqrt{-\log(1-1/n)}$ and $\delta$ as $2\zeta/k$.
\subsection{Evaluation of Efficiency}\label{subsec:runtime}
In this section we first compare the proposed global BCGD $\texttt{BCGD}_G$,  local BCGD $\texttt{BCGD}_L$, and incremental BCGD $\texttt{BCGD}_I$ with other latent space inferring approaches in terms of inference time and memory consumption. We then subsequently compare BCGD approaches with two auto-dimensionality latent space approaches BIGCLAIM and LabelRT and graph heuristic AA in terms of online link prediction time and memory consumption.
\begin{figure}[!t]
  \centering
  \includegraphics[width=0.7\columnwidth]{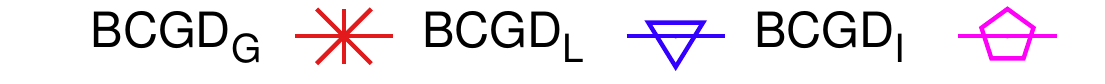}
  \subfigure[Inferring time (s)]{\includegraphics[width=0.492\columnwidth]{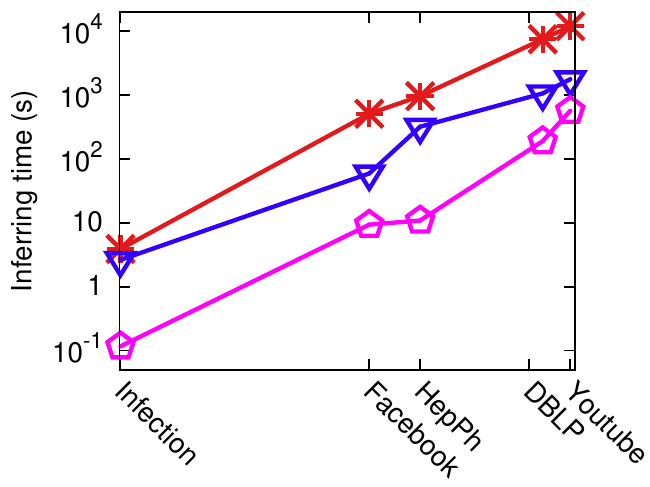}}
  \subfigure[Memory usage (MB)]{\includegraphics[width=0.492\columnwidth]{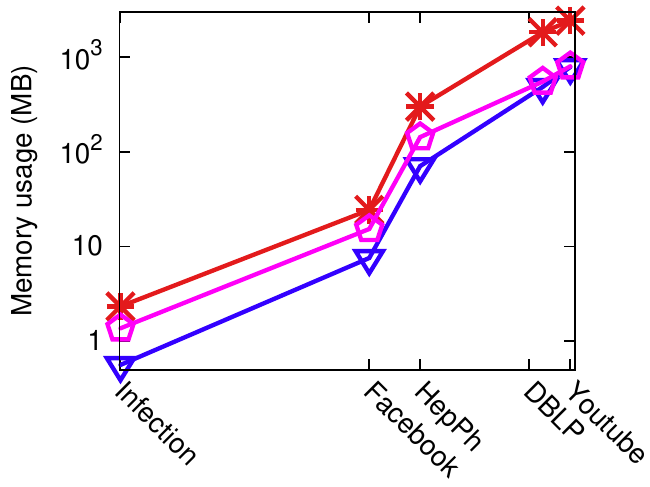}}
  \caption{\small{Inference efficiency of BCGD algorithms}}\label{fig:bcgdtime}
\end{figure}

\vspace{0.1cm}
\noindent \textbf{Offline inference efficiency.} We first report the total running time and memory consumption of our BCGD approaches in Figure~\ref{fig:bcgdtime}. Clearly, $\texttt{BCGD}_G$ is neither memory efficient nor time efficient. Both $\texttt{BCGD}_I$ and $\texttt{BCGD}_L$ are very time and memory efficient: $\texttt{BCGD}_I$ is  more efficient in running time, while $\texttt{BCGD}_L$ is more efficient in memory usage. This is because $\texttt{BCGD}_I$ requires additional storage for graph changes $\Delta G$; but it adaptively updates latent space with the graph changes and thus is very time efficient.

\begin{figure}[thb]
\centering
 \includegraphics[width=\columnwidth]{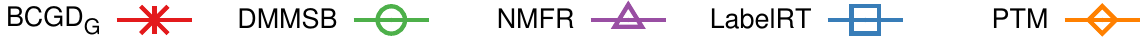}
\subfigure[Inferring time (s)]{\label{fig:timesmall}\includegraphics[width=0.492\columnwidth]{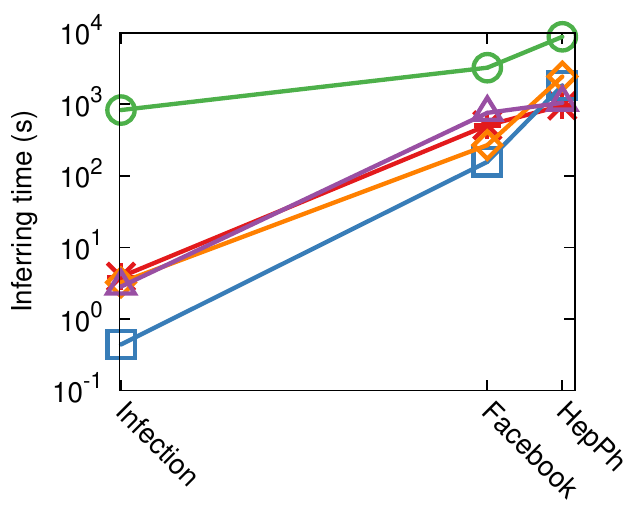}}
\subfigure[Memory usage (MB)]{\label{fig:memorysmall}\includegraphics[width=0.492\columnwidth]{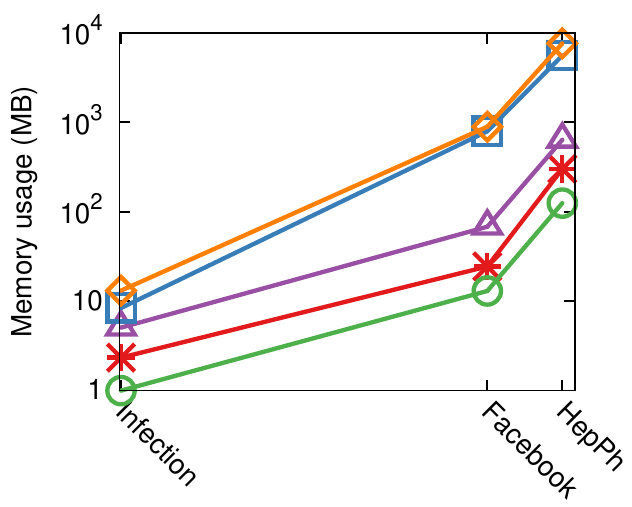}}
\vspace{-4mm}
\caption{\small{Inference efficiency comparison with DMMSB, PTM, LabelRT and NMFR.}}\label{fig:effcomp1}
\end{figure}

In the following we group other baselines into two classes to facilitate results visualization. We first compare those memory- or time-inefficient approaches DMMSB, PTM, LabelRT and NMFR with our $\texttt{BCGD}_G$ and report the results in Figure~\ref{fig:effcomp1}. Among them, PTM and LabelRT require much larger memory consumption and thus fail to process the two large graph DBLP and YouTube due to their memory bottlenecks. NMFR is more memory efficient than PTM and LabelRT, but it still can not handle the two large graphs. In addition, global optimization algorithm DMMSB takes too long to process the two large graph DBLP and Youtube due to running time bottleneck. Figure~\ref{fig:timesmall} shows that on Infection and Facebook data, LabelRT $\gg_t$ PTM $\gg_t$ \{NMFR, $\texttt{BCGD}_G$\} $\gg_t$ DMMSB, where $ A\gg_t B$ denotes that A is faster than B on average. On HepPh data, \{NMFR, $\texttt{BCGD}_G$\} $\gg_t$ \{LabelRT, PTM\} $\gg_t$ DMMSB. Clearly, our global optimization algorithm $\texttt{BCGD}_G$ is at least five times faster than the other global approach, DMMSB, and is comparable to inefficient local approach NMFR. For small graphs, it is not surprising that the most efficient algorithm is the incremental approach LabelRT since it incrementally maintains and updates latent spaces. Unfortunately, it consumes large amounts of memory, and becomes much slower than our approach $\texttt{BCGD}_G$ for graph HepPh. In addition, all of these approaches are much slower than our fast BCGD algorithms $\texttt{BCGD}_L$ and $\texttt{BCGD}_I$, each of which takes less than 400 seconds to finish learning temporal latent space for HepPh.

\begin{figure}[thb]
\centering
\includegraphics[width=\columnwidth]{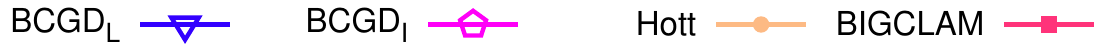}
\subfigure[Inferring time (s)]{\label{fig:timelaege}\includegraphics[width=0.492\columnwidth]{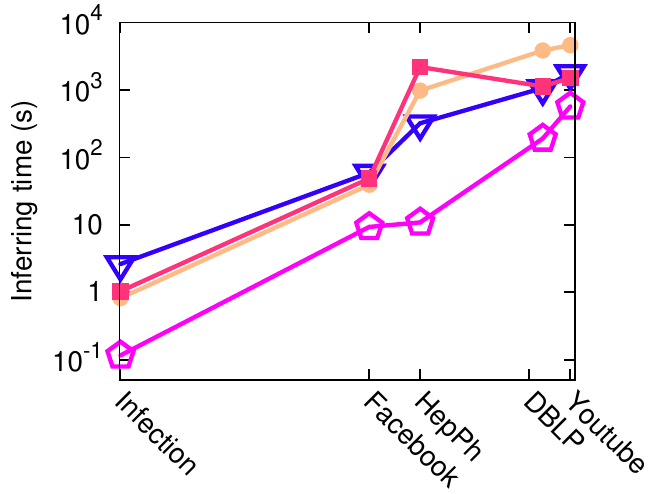}}
\subfigure[Memory usage (MB)]{\label{fig:memorylarge}\includegraphics[width=0.492\columnwidth]{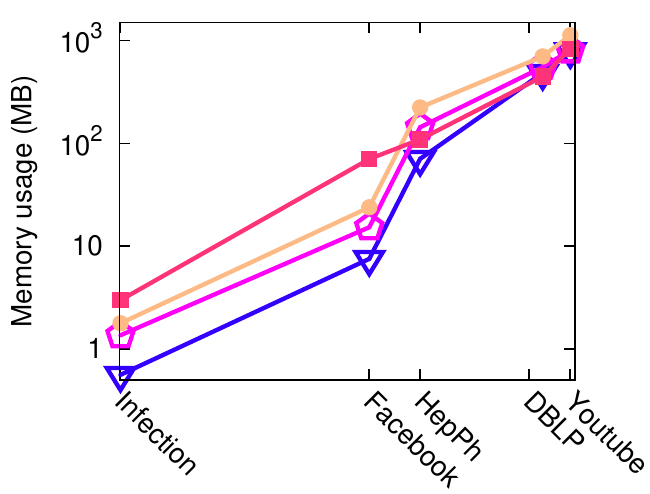}}
\caption{\small{Inference efficiency comparison with scalable baselines Hott and BIGCLAM.}}\label{fig:effcomp2}
\end{figure}

We further compare our fast BCGD algorithms with state-of-the-arts scalable approaches Hott and BigCLAM. $\texttt{BCGD}_L$ is comparable to Hott and BigCLAM. In addition, the incremental approach $\texttt{BCGD}_I$ is consistently more efficient than alternative approaches Hott and BigCLAM. Because we noticed from Figure~\ref{fig:memorylarge} that the memory usage of these four approaches is similar, we conclude that our two fast BCGD algorithms are both memory- and time-efficient.

\begin{figure}[thb]
\centering
\includegraphics[width=\columnwidth]{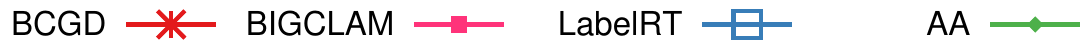}
\subfigure[Prediction time (s)]{\label{fig:Ptime}\includegraphics[width=0.492\columnwidth]{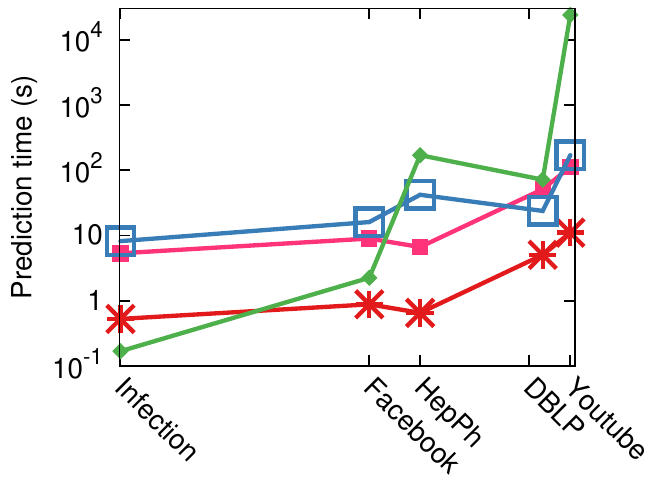}}
\subfigure[Memory usage (MB)]{\label{fig:Pmemory}\includegraphics[width=0.492\columnwidth]{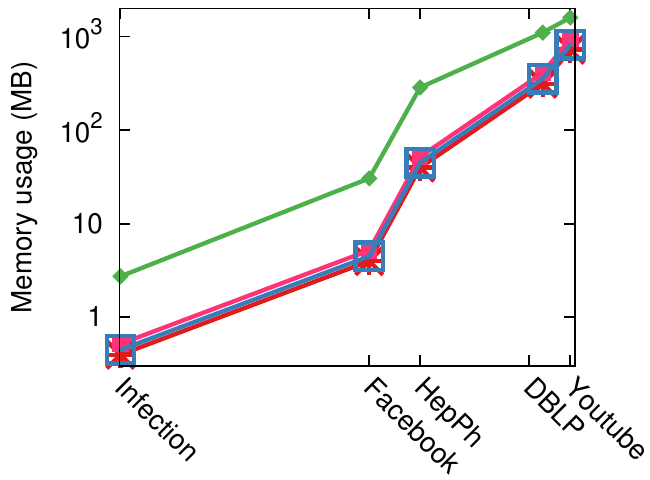}}
\caption{\small{Online prediction time and memory comparison for 20 thousand  node pairs.} Note that BCGD represents all of the three proposed algorithms ($\texttt{BCGD}_G$, $\texttt{BCGD}_L$, and $\texttt{BCGD}_I$) since the online prediction time and memory consumption of these three are the same.}\label{fig:predictiontime}
\vspace{-0.4cm}
\end{figure}

\vspace{0.1cm}
\noindent \textbf{Online Prediction Efficiency.} We now verify that the low-dimension latent space approach is very efficient for online link prediction. Here we choose AA as a representative graph heuristic approach for link predication, not only because it achieves good accuracy~\cite{SarkarCJICML12}, but also because it is fast to compute compared to other graph heuristics~\cite{Cohen2013}. However, AA is still much slower than latent space-based approaches in online link probability computation. As shown in Figure~\ref{fig:predictiontime}, on the HepPh dataset, for 20,000 node pairs, it takes more than 150 seconds to compute the AA scores; while all the low-dimension latent space-based approaches are able to finish online link probability computation in less than one second. In addition, latent space-based link prediction approaches are also more efficient than AA in terms of memory consumption, see Figure~\ref{fig:Pmemory}. To predict whether two nodes are linked, latent space-based approaches only need to read the latent positions of two nodes; while AA requires the neighborhood information of two nodes. Finally, Figure~\ref{fig:Ptime} also intuitively supports the reason why we use low-dimension constraint. With low-dimension constraint ($k$=20), our BCGD approach obtains at least four times speed-up than auto-dimensionality approaches LabelRT and BigCLAM (sometimes $k$ can be more than two hundred) in online link prediction.

\subsection{Evaluation of Link Prediction Accuracy}
\begin{figure}[!t]
\centering
\label{fig:ROCsmall1}\includegraphics[width=0.9\columnwidth]{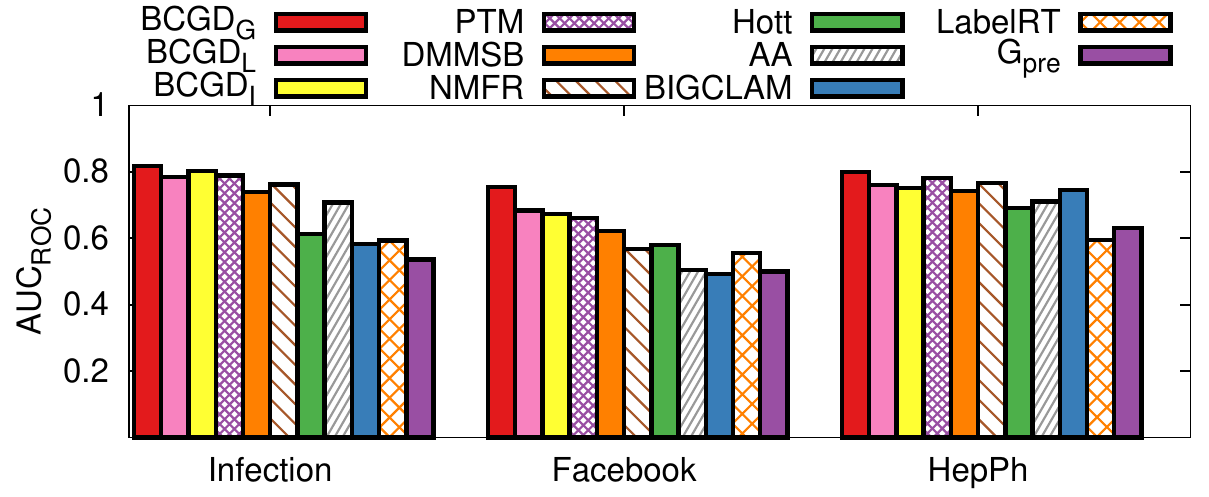}
\label{fig:ROClarge1}\includegraphics[width=0.9\columnwidth]{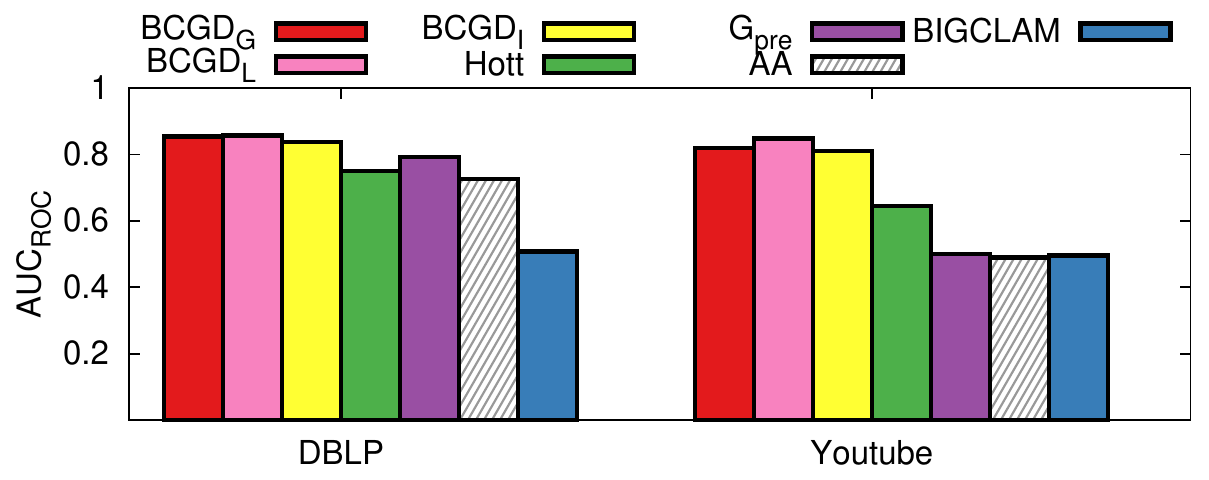}
\caption{\small{All links prediction accuracy comparison.}}\label{fig:Accall}
\vspace{-0.3cm}
\end{figure}
In this section we quantitatively evaluate the quality of learned latent spaces in terms of their predictive power. We give an overall comparison of all the approaches in terms of $AUC_{ROC}$ on predicting all links and report the results in Figure~\ref{fig:Accall}. Here, $G_{\texttt{pre}}$ denotes the simple baseline using $G_t$ to predict $G_{t+1}$. Clearly, for all link prediction accuracy, we have \{$\texttt{BCGD}_G$, $\texttt{BCGD}_L$, $\texttt{BCGD}_I$, PTM, DMMSB\} $\gg_a$ \{Hott, NMFR\} $\gg_a$ \{BIGCLAM, LabelRT, AA, $G_{\texttt{pre}}$\}, where $A \gg_a B$ denotes that on average,  $A$ is more accurate than $B$ in terms of predictive power.  BigCLAM and LabelRT have poor AUC scores because they only output a single, hard assignment of nodes to each dimension, which provides no means to control the trade-off between true and false positive rates. AA did not perform well on Facebook and YouTube because it can not capture the deleted links prediction; AA still gives each unlinked node pairs a high score based on their topological similarity on aggregated graph from $G_1$ to $G_t$. Our methods perform much better than $G_{pre}$ due to: 1) $Z_t$ encodes not only the link information of $G_t$, but also the temporal information from $G_1$ to $G_t$; 2) $Z_t$ is a good estimate for $Z_{t+1}$; 3) the reconstruction error between $Z_tZ_t^T$ and $G_t$ reflects the possible future changes. Given any pair $(u, v)$, if the reconstruction error between $Z_t(u)Z_t(v)^T$ and $G_t(u, v)$ is large, then it indicates that the proximity of the pair $(u, v)$ is changing over time.

\begin{figure}[!t]
\centering
\label{fig:ROCsmall2}\includegraphics[width=0.9\columnwidth]{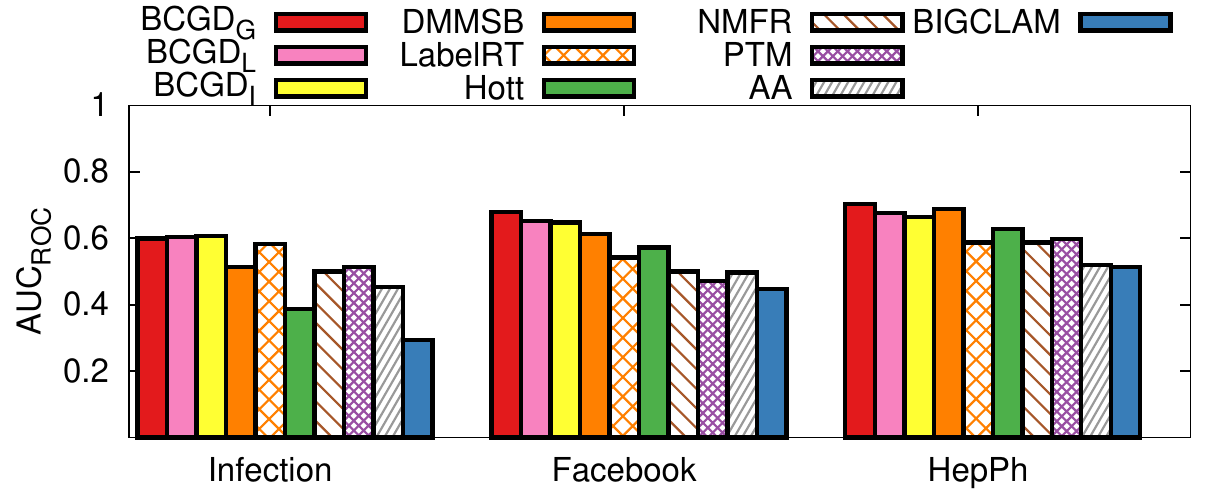}
\label{fig:ROClarge2}\includegraphics[width=0.9\columnwidth]{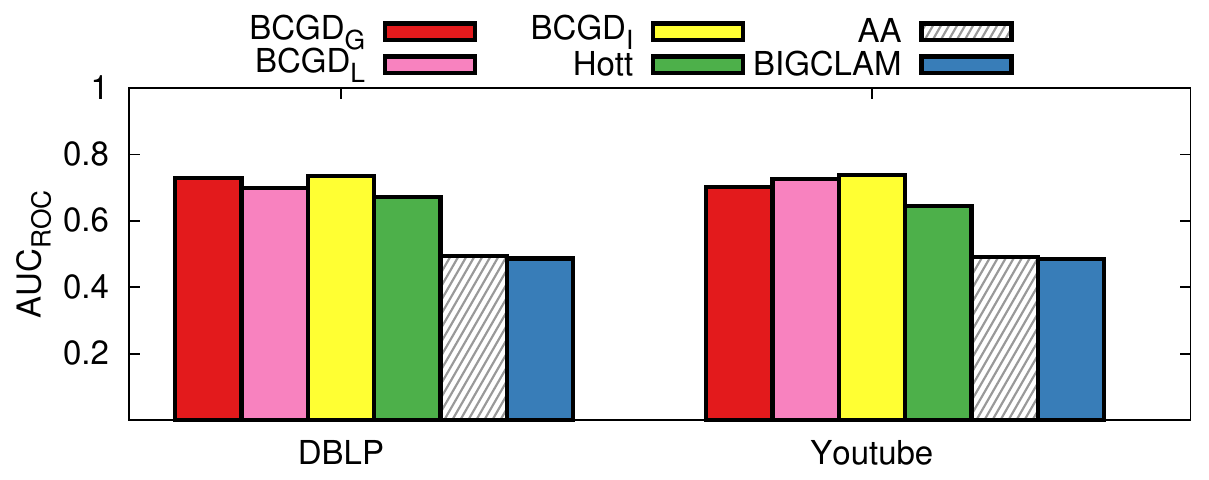}
\caption{\small{New links prediction accuracy comparison.}}\label{fig:Accnew}
\vspace{-0.3cm}
\end{figure}

We next evaluate the link prediction performance of all the approaches on new links. The results are plotted in Figure~\ref{fig:Accnew}. As expected, the AUC values are much lower for the new links. For the new links prediction, we have \{$\texttt{BCGD}_G$, $\texttt{BCGD}_L$, $\texttt{BCGD}_I$, LabelRT, DMMSB\} $\gg_a$ \{Hott, NMFR\} $\gg_a$ \{AA, BIGCLAM\}. It is not surprising that all the temporal link prediction approaches perform better than static link prediction approaches.

Figure~\ref{fig:Accnew} also reflects an interesting phenomenon: our fast BCGD algorithms $\texttt{BCGD}_L$ and $\texttt{BCGD}_I$ not only improve the efficiency bottleneck of latent space inference, but also achieve comparable prediction accuracy with the global optimization approach $\texttt{BCGD}_G$. For the new links prediction, they even outperform global approaches on both dataset Infection and YouTube. This is because both $\texttt{BCGD}_L$ and $\texttt{BCGD}_I$ utilize local structures and local invariance, which significantly enhances local and temporal information encoding in the temporal latent space. In addition, we notice that on average, the incremental approach $\texttt{BCGD}_I$ is able to obtain higher accuracy than the local approach $\texttt{BCGD}_L$. Although both  $\texttt{BCGD}_L$ and $\texttt{BCGD}_I$ utilize local invariance, $\texttt{BCGD}_I$ is able to exploit the local structure and local invariance in an adaptive way with graph changes, which leads to a better performance. Because $\texttt{BCGD}_I$ is more scalable than all the other alternative approaches (as shown in Figure~\ref{fig:timelaege}). We conclude that $\texttt{BCGD}_I$ is a practical solution for link prediction on large-scale dynamic networks.

\section{Related work}\label{sec:related}
Recently, link prediction has attracted significant attention, and various methods have been proposed for different types of networks. In the following, we first give a brief overview of related work in link prediction. Particularly, we focus on two categories: graph-based heuristics and latent space-based approaches. Next, we present some related work in inferring latent positions of nodes in a latent space.
\subsection{Link Prediction}
The link prediction problem in static networks has been extensively studied. We refer readers to two recent survey papers~\cite{NowellCIKM2003,Mohammadsurveylinkprediction} and the related work section of the paper~\cite{Zhu14socialcom} for an exhaustive introduction to this thriving field. Among these existing works, one of the most popular categories of methods is graph-based heuristics. Graph-based heuristics~\cite{NowellCIKM2003,Adamic01friendsand,Symeonidis2010,LeoPsy1953,Jehsimrank2002} model the link probability between two nodes as a function of their topological similarity. The topological similarity can be defined based on local proximity such as common neighbors~\cite{Adamic01friendsand}, triad closures~\cite{Symeonidis2010}, or global proximity such as weighted shortest-paths~\cite{LeoPsy1953}. Here we highlight Adamic and Adar (\emph{AA})~\cite{Adamic01friendsand} who proposed a degree-weighted common-neighbor heuristic that works well in collaboration networks.

Unfortunately, in online link prediction, the computation of topological similarity metrics, especially for those metrics that require path computation or random walk over the entire network, is time-consuming if computed on-the-fly~\cite{Cohen2013}. On the other hand, one can precompute all pairs of topological similarity scores in advance, leading to a quadratic storage cost. To strike a balance between online computation cost and offline storage cost, recent research interests in link prediction have been directed towards temporal link prediction with latent space modeling~\cite{Hoff2002}. For example, Fu et al.~\cite{Fu:2009:DMM:1553374.1553416} extended the mixed membership block model to allow a linear Gaussian trend in the model parameters (\emph{DMMSB}). Sewell and Chen~\cite{sewell2015latent} proposed a model which embeds longitudinal network data as trajectories in a latent Euclidean space, and the probability of link formation between two nodes depends on their Euclidean distance and popularity. In this work we do not take the popularity of each node into consideration since we assume that the popularity of each node is automatically captured by the latent space modeling: a popular node is surrounded by many other nodes in the latent space. Furthermore, our model penalizes sudden and large changes in the latent positions.

Dunlavy et al.~\cite{DunlavyTKDDTLP} developed a tensor-based latent space modeling to predict temporal links. Unfortunately, in online prediction, it requires the tensor product of vectors, which is much more computationally expensive than the dot product of vectors in the proposed approach. In addition, their approaches require very large storage costs since they need to put the entire tensor into memory. Recently, there have been several more efficient approaches to conduct tensor decomposition. For instance, Huang et al.~\cite{HuangNHVA13} proposed to conduct tensor decomposition with singular value decomposition to find temporal communities in a GPGPU setting.


\subsection{Inferring Latent Space}
Recent work has explored static latent space modeling for social network analysis. For example, Hoff et al.~\cite{Hoff2002} developed a class of models where the probability of a relation between actors depends on the positions of individuals in an unobserved social space. They make inference for the latent space with the maximum likelihood estimation and Bayesian posterior inference. Obtaining this posterior is, however, often infeasible in large-scale graphs. Yin et al.~\cite{YinHX13} proposed an efficient stochastic variational inference algorithm and a parsimonious triangular model to infer the latent spaces of large networks (\emph{PTM}). Sarkar and Moore~\cite{Sarkarlatentspace2005} first generalized the classical multidimensional scaling to get an initial estimate of the positions in the latent space, and then applied nonlinear optimization directly to obtain the latent space representation.

Matrix factorization approaches are also applied to observed networks to learn the low-dimension latent space representation.  Yang et al.~\cite{YangnIPS12} propose a multiplicative algorithm with graph random walk to solve the symmetric graph factorization problem (\emph{NMFR}). However, their approach does not scale well due to the high computation cost in each iteration. Additionally, the effectiveness of their approach decreases as the density of the input graph increases. Yang and Leskovec~\cite{YangWSDM2013} proposed a matrix factorization approach over networks to learn the static latent space representation by maximizing the link likelihood (\emph{BIGCLAM}). Their approach is very scalable; nevertheless, the learned latent space is of high dimension, which leads to expensive online link prediction computation.  HottTopixx (\emph{Hott}~\cite{RechtRTBNIPS12}) uses a new approach for NMF with low-rank constraint which is highly parallel and allows it to run on very large datasets. Unfortunately, their approach is designed to factorize the traditional rectangle matrices, and scalable approaches in symmetric graph factorization are much less studied than rectangle matrix factorization such as Hott. In this work we apply symmetric matrix factorization approaches directly on the observed networks to infer low-rank latent spaces. We propose a block coordinate gradient descent algorithm, which is more scalable than NMFR and performs well, regardless of the topology of input graphs.

In addition to matrix factorization approaches, recently Xie et al.~\cite{Xiecorr} (LabelRT) proposed a label propagation-based approach that incrementally detects communities in dynamic networks. In this work we propose an incremental BCGD algorithm to incrementally detect latent positions at each time step with conditional updates on a set of affected nodes.

\section{Conclusion and Future Work}\label{sec:con}
In this work we propose a scalable approach for link prediction with a temporal latent space model, which assumes two nodes are more likely to be linked if they are close to each other in their latent space. 
In addition, our temporal latent space model prefers smoothly evolving by penalizing frequent changes in latent positions for each node. With respect to the proposed model, we developed three different inference algorithms, $\texttt{BCGD}_G$, $\texttt{BCGD}_L$ and $\texttt{BCGD}_I$ to learn the temporal latent space via non-negative matrix factorization approaches. We also provide a set of theoretical analyses characterizing their performance guarantees. We conducted a set of experiments on large networks with millions of nodes and links to verify both the efficiency and predictive quality of all the proposed approaches.

Our model still has limitations. First, the temporal smoothness assumption may not hold in some circumstance. For example, in a road network, the latent position of each node changes significantly from rush hour to a non-rush hour. And even in social networks,  where temporal smoothness assumptions typically hold, external events may cause significant shifts to the network that reflect abrupt changes in latent node positions. We plan to extend the proposed model to support temporal nonlinear transitions. Second, we plan to propose a new continuous time model that supports continuous inputs rather than discretized graph snapshots. Moreover, we plan to extend our approach to generalized cases including directed and weighted graphs. In addition to model improvement, another interesting direction is to further improve the efficiency. Note that in our block coordinate gradient descent approach, the latent position update for a node $u$ can be conducted simultaneously with another node $v$ if they do not share the same neighborhood. This property can be leveraged for very good parallelization in the future.

\section*{Acknowledgment}
We are very grateful to Prof. Dacheng Tao, Prof. Kun Zhang, Prof. Rong Ge, Li Han and Dingxiong Deng for their insightful discussions. This research was supported in part by DARPA grant No. W911NF–12–1–0034.
\bibliographystyle{IEEEtranS}
\bibliography{socialtie}

\begin{thebibliography}{10}
\providecommand{\url}[1]{#1}
\csname url@samestyle\endcsname
\providecommand{\newblock}{\relax}
\providecommand{\bibinfo}[2]{#2}
\providecommand{\BIBentrySTDinterwordspacing}{\spaceskip=0pt\relax}
\providecommand{\BIBentryALTinterwordstretchfactor}{4}
\providecommand{\BIBentryALTinterwordspacing}{\spaceskip=\fontdimen2\font plus
\BIBentryALTinterwordstretchfactor\fontdimen3\font minus
  \fontdimen4\font\relax}
\providecommand{\BIBforeignlanguage}[2]{{%
\expandafter\ifx\csname l@#1\endcsname\relax
\typeout{** WARNING: IEEEtranS.bst: No hyphenation pattern has been}%
\typeout{** loaded for the language `#1'. Using the pattern for}%
\typeout{** the default language instead.}%
\else
\language=\csname l@#1\endcsname
\fi
#2}}
\providecommand{\BIBdecl}{\relax}
\BIBdecl

\bibitem{konect:2015:dblp_coauthor}
\BIBentryALTinterwordspacing
``Dblp network dataset -- {KONECT},'' May 2015. [Online]. Available:
  \url{http://konect.uni-koblenz.de/networks/dblp_coauthor}
\BIBentrySTDinterwordspacing

\bibitem{Adamic01friendsand}
L.~A. Adamic and E.~Adar, ``Friends and neighbors on the web,'' \emph{SOCIAL
  NETWORKS}, vol.~25, pp. 211--230, 2001.

\bibitem{Airoldi:2008:MMS:1390681.1442798}
E.~M. Airoldi, D.~M. Blei, S.~E. Fienberg, and E.~P. Xing, ``Mixed membership
  stochastic blockmodels,'' \emph{J. Mach. Learn. Res.}, vol.~9, pp.
  1981--2014, 2008.

\bibitem{AroraSTOC2012}
S.~Arora, R.~Ge, R.~Kannan, and A.~Moitra, ``Computing a nonnegative matrix
  factorization -- provably,'' in \emph{STOC Conference}, 2012, pp. 145--162.

\bibitem{beck2009fast}
A.~Beck and M.~Teboulle, ``A fast iterative shrinkage-thresholding algorithm
  for linear inverse problems,'' \emph{SIAM journal on imaging sciences},
  vol.~2, no.~1, pp. 183--202, 2009.

\bibitem{Berry2007}
M.~W. Berry, M.~Browne, A.~N. Langville, V.~P. Pauca, and R.~J. Plemmons,
  ``Algorithms and applications for approximate nonnegative matrix
  factorization,'' \emph{Comp. Stat. \& Data Analysis}, vol.~52, no.~1, pp.
  155--173, 2007.

\bibitem{Cohen2013}
S.~Cohen and N.~Cohen-Tzemach, ``Implementing link-prediction for social
  networks in a database system,'' in \emph{SIGMOD DBSocial Workshop}, 2013,
  pp. 37--42.

\bibitem{DunlavyTKDDTLP}
D.~M. Dunlavy, T.~G. Kolda, and E.~Acar, ``Temporal link prediction using
  matrix and tensor factorizations,'' \emph{ACM Trans. Knowl. Discov. Data},
  vol.~5, no.~2, pp. 10:1--10:27, 2011.

\bibitem{ErdosTKDD2014}
D.~Erd\H{o}s, R.~Gemulla, and E.~Terzi, ``Reconstructing graphs from
  neighborhood data,'' \emph{ACM Trans. Knowl. Discov. Data}, vol.~8, no.~4,
  pp. 23:1--23:22, 2014.

\bibitem{Fu:2009:DMM:1553374.1553416}
W.~Fu, L.~Song, and E.~P. Xing, ``Dynamic mixed membership blockmodel for
  evolving networks,'' in \emph{ICML Conference}, 2009, pp. 329--336.

\bibitem{GaoCIKM2011TLP}
S.~Gao, L.~Denoyer, and P.~Gallinari, ``Temporal link prediction by integrating
  content and structure information,'' in \emph{CIKM Conference}, 2011, pp.
  1169--1174.

\bibitem{Gibson2005}
D.~Gibson, R.~Kumar, and A.~Tomkins, ``Discovering large dense subgraphs in
  massive graphs,'' in \emph{VLDB Conference}, 2005, pp. 721--732.

\bibitem{journals/tsp/GuanTLY12}
N.~Guan, D.~Tao, Z.~Luo, and B.~Yuan, ``Nenmf: An optimal gradient method for
  nonnegative matrix factorization.'' \emph{IEEE Trans. on Signal Processing},
  pp. 2882--2898, 2012.

\bibitem{VLDBGupta2014}
P.~Gupta, V.~Satuluri, A.~Grewal, S.~Gurumurthy, V.~Zhabiuk, Q.~Li, and J.~Lin,
  ``Real-time twitter recommendation: Online motif detection in large dynamic
  graphs,'' \emph{Proc. VLDB Endow.}, vol.~7, no.~13, pp. 1379--1380, 2014.

\bibitem{Mohammadsurveylinkprediction}
M.~A. Hasan and M.~J. Zaki, ``A survey of link prediction in social networks,''
  in \emph{Social Network Data Analytics}.\hskip 1em plus 0.5em minus
  0.4em\relax Springer US, 2011, pp. 243--275.

\bibitem{Hoff2002}
P.~D. Hoff, A.~E. Raftery, and M.~S. Handcock, ``Latent space approaches to
  social network analysis,'' \emph{Journal of the American Statistical
  Association}, vol.~97, 2002.

\bibitem{HuangNHVA13}
F.~Huang, N.~U. N, M.~U. Hakeem, P.~Verma, and A.~Anandkumar, ``Fast detection
  of overlapping communities via online tensor methods on gpus.'' \emph{CoRR},
  vol. abs/1309.0787, 2013.

\bibitem{konect:sociopatterns}
L.~Isella, J.~Stehl\'{e}, A.~Barrat, C.~Cattuto, J.-F. Pinton, and W.~V. den
  Broeck, ``What's in a crowd? analysis of face-to-face behavioral networks,''
  \emph{J. of Theoretical Biology}, vol. 271, no.~1, pp. 166--180, 2011.

\bibitem{Jehsimrank2002}
G.~Jeh and J.~Widom, ``Simrank: a measure of structural-context similarity,''
  in \emph{SIGKDD Conference}, 2002, pp. 538--543.

\bibitem{LeoPsy1953}
L.~Katz, ``A new status index derived from sociometric analysis,''
  \emph{Psychometrika}, vol.~18, no.~1, pp. 39--43, March 1953.

\bibitem{KONECT}
J.~Kunegis, ``{KONECT - The Koblenz Network Collection},'' in \emph{Proc. Int.
  Web Observatory Workshop}, 2013.

\bibitem{lee99}
D.~D. Lee and H.~S. Seung, ``Learning the parts of objects by nonnegative
  matrix factorization,'' \emph{Nature}, vol. 401, no. 6755, pp. 788--791,
  1999.

\bibitem{konect:leskovec107}
J.~Leskovec, J.~Kleinberg, and C.~Faloutsos, ``Graph evolution: Densification
  and shrinking diameters,'' \emph{ACM Trans. Knowledge Discovery from Data},
  vol.~1, no.~1, pp. 1--40, 2007.

\bibitem{NowellCIKM2003}
D.~Liben-Nowell and J.~Kleinberg, ``The link prediction problem for social
  networks,'' in \emph{CIKM Conference}, 2003, pp. 556--559.

\bibitem{homophily}
M.~Mcpherson, L.~Smith-Lovin, and J.~M. Cook, ``Birds of a feather: Homophily
  in social networks,'' \emph{Annual Review of Sociology}, vol.~27, pp.
  415--444, 2001.

\bibitem{Menon:2011:LPV:2034117.2034146}
A.~K. Menon and C.~Elkan, ``Link prediction via matrix factorization,'' in
  \emph{ECML/PKDD Conference}, 2011, pp. 437--452.

\bibitem{konect:mislove2}
A.~Mislove, ``Online social networks: Measurement, analysis, and applications
  to distributed information systems,'' Ph.D. dissertation, Department of
  Computer Science, Rice University, 2009.

\bibitem{opac-b1104789}
Y.~Nesterov, \emph{Introductory lectures on convex optimization : a basic
  course}.\hskip 1em plus 0.5em minus 0.4em\relax Kluwer Academic Publ., 2004.

\bibitem{QiICDE2013}
G.-J. Qi, C.~C. Aggarwal, and T.~Huang, ``Link prediction across networks by
  biased cross-network sampling,'' in \emph{ICDE Conference}, 2013, pp.
  793--804.

\bibitem{RechtRTBNIPS12}
B.~Recht, C.~Re, J.~A. Tropp, and V.~Bittorf, ``Factoring nonnegative matrices
  with linear programs,'' in \emph{NIPS Conference}, 2012, pp. 1223--1231.

\bibitem{SarkarCJICML12}
P.~Sarkar, D.~Chakrabarti, and M.~I. Jordan, ``Nonparametric link prediction in
  dynamic networks,'' in \emph{ICML Conference}, 2012, pp. 1687--1694.

\bibitem{Sarkarlatentspace2005}
P.~Sarkar and A.~W. Moore, ``Dynamic social network analysis using latent space
  models,'' \emph{SIGKDD Explor. Newsl.}, vol.~7, no.~2, pp. 31--40, 2005.

\bibitem{sewell2015latent}
D.~K. Sewell and Y.~Chen, ``Latent space models for dynamic networks,''
  \emph{Journal of the American Statistical Association}, vol. 110, no. 512,
  pp. 1646--1657, 2015.

\bibitem{Sun2007}
J.~Sun, C.~Faloutsos, S.~Papadimitriou, and P.~S. Yu, ``Graphscope:
  Parameter-free mining of large time-evolving graphs,'' in \emph{SIGKDD
  Conference}, 2007, pp. 687--696.

\bibitem{Symeonidis2010}
P.~Symeonidis, E.~Tiakas, and Y.~Manolopoulos, ``Transitive node similarity for
  link prediction in social networks with positive and negative links,'' in
  \emph{Proceedings of the 4th ACM conference on Recommender systems}, 2010,
  pp. 183--190.

\bibitem{TantipathananandhKDD2007}
C.~Tantipathananandh, T.~Berger-Wolf, and D.~Kempe, ``A framework for community
  identification in dynamic social networks,'' in \emph{SIGKDD Conference},
  2007, pp. 717--726.

\bibitem{Tantipathananandh:2011:FCD:2117684.2118320}
C.~Tantipathananandh and T.~Y. Berger-Wolf, ``Finding communities in dynamic
  social networks,'' in \emph{ICDM Conference}, 2011, pp. 1236--1241.

\bibitem{TsengY09}
P.~Tseng and S.~Yun, ``A coordinate gradient descent method for nonsmooth
  separable minimization,'' \emph{Math. Program.}, vol. 117, no. 1-2, pp.
  387--423, 2009.

\bibitem{Vavasis2009}
S.~A. Vavasis, ``On the complexity of nonnegative matrix factorization,''
  \emph{J. on Optimization}, vol.~20, no.~3, pp. 1364--1377, 2009.

\bibitem{viswanath09}
B.~Viswanath, A.~Mislove, M.~Cha, and K.~P. Gummadi, ``On the evolution of user
  interaction in {Facebook},'' in \emph{Proc. Workshop on Online Social
  Networks}, 2009, pp. 37--42.

\bibitem{Xiecorr}
J.~Xie, M.~Chen, and B.~K. Szymanski, ``Labelrankt: Incremental community
  detection in dynamic networks via label propagation,'' \emph{CoRR}, vol.
  abs/1305.2006, 2013.

\bibitem{YangWSDM2013}
J.~Yang and J.~Leskovec, ``Overlapping community detection at scale: A
  nonnegative matrix factorization approach,'' in \emph{WSDM Conference}, 2013,
  pp. 587--596.

\bibitem{YangnIPS12}
Z.~Yang, T.~Hao, O.~Dikmen, X.~Chen, and E.~Oja, ``Clustering by nonnegative
  matrix factorization using graph random walk,'' in \emph{NIPS Conference},
  2012, pp. 1088--1096.

\bibitem{YeWWW2013}
J.~Ye, H.~Cheng, Z.~Zhu, and M.~Chen, ``Predicting positive and negative links
  in signed social networks by transfer learning,'' in \emph{WWW Conference},
  2013, pp. 1477--1488.

\bibitem{YinHX13}
J.~Yin, Q.~Ho, and E.~P. Xing, ``A scalable approach to probabilistic latent
  space inference of large-scale networks,'' in \emph{NIPS Conference}, 2013,
  pp. 422--430.

\bibitem{Zhang:2014ICD}
J.~Zhang, C.~Wang, J.~Wang, and J.~X. Yu, ``Inferring continuous dynamic social
  influence and personal preference for temporal behavior prediction,''
  \emph{Proc. VLDB Endow.}, vol.~8, no.~3, pp. 269--280, 2014.

\bibitem{WWW13}
Y.~Zhang, M.~Zhang, Y.~Liu, S.~Ma, and S.~Feng, ``Localized matrix
  factorization for recommendation based on matrix block diagonal forms,'' in
  \emph{WWW Conference}, 2013, pp. 1511--1520.

\bibitem{ZhuGCL14}
L.~Zhu, A.~Galstyan, J.~Cheng, and K.~Lerman, ``Tripartite graph clustering for
  dynamic sentiment analysis on social media,'' in \emph{SIGMOD Conference},
  2014, pp. 1531--1542.

\bibitem{Zhu14socialcom}
L.~Zhu and K.~Lerman, ``A visibility-based model for link prediction in social
  media,'' in \emph{ASE SocialCom Conference}, 2014.

\end{thebibliography}
\begin{IEEEbiography}[{\includegraphics[width=1in,height=1.25in,clip,keepaspectratio]{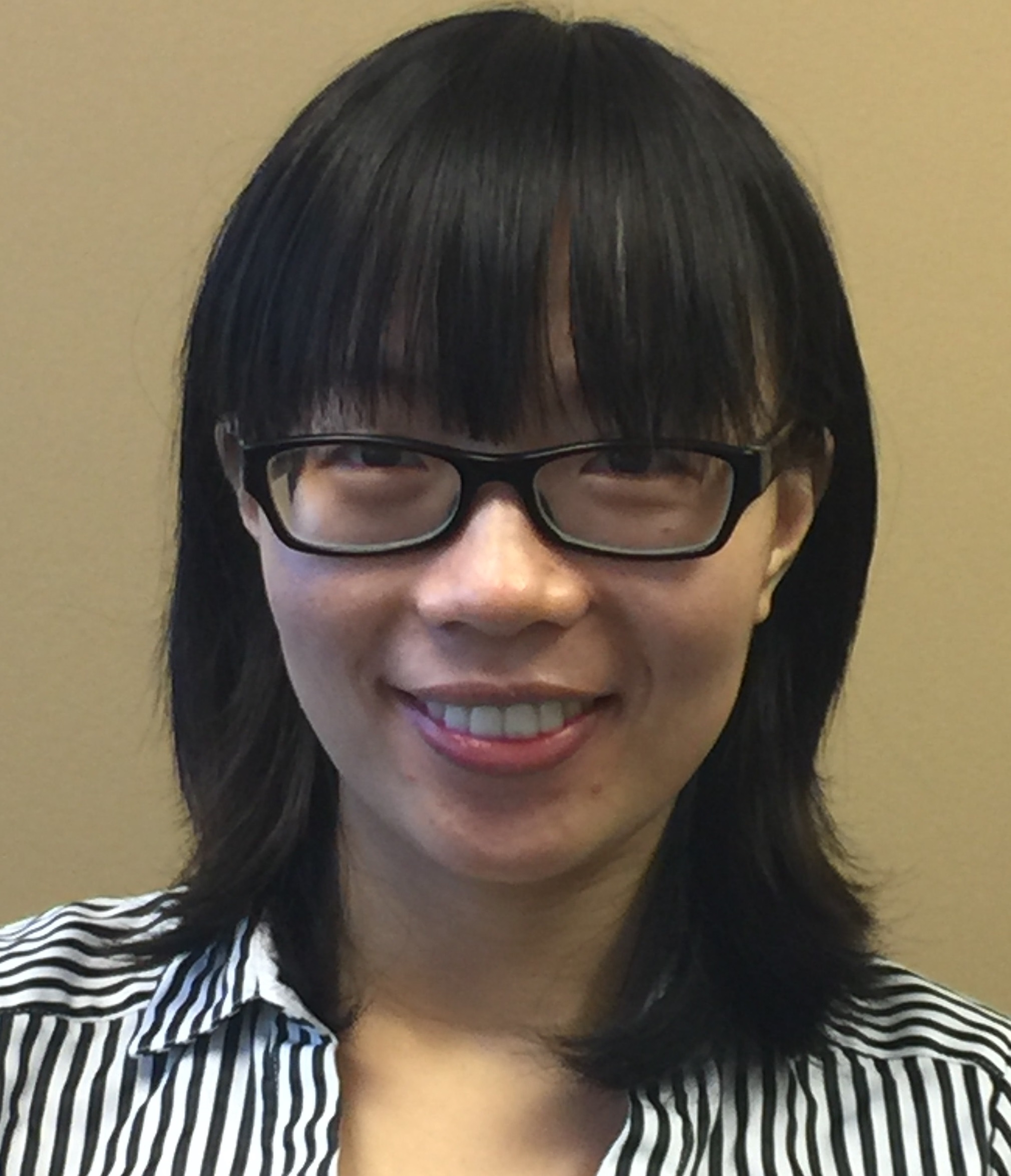}}]{Linhong Zhu}
is a computer scientist at USC's Information Sciences Institute. Prior to that she worked as a Scientist-I in the data analytics department at the Institute for Infocomm Research, Singapore. She obtained her Ph.D. degree in computer engineering from Nanyang Technological University, Singapore in 2011 and the BS degree from the University of Science and Technology of
China in 2006. Her research interests are in large-scale graph analytics with applications to social network analysis, social media analysis, and predictive modeling. She is a member of IEEE and ACM.
\end{IEEEbiography}
\vspace{-0.8cm}
\begin{IEEEbiography}[{\includegraphics[width=1in,height=1.25in,clip,keepaspectratio]{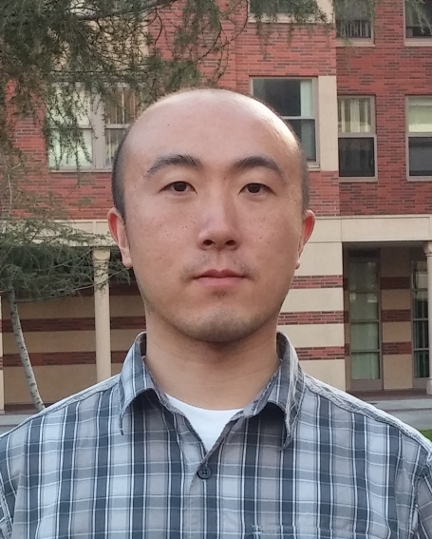}}]{Dong Guo}
received the BS degree in physics from the University of Science and Technology of China, Hefei, China, in 2007, and entered the Ph.D. program in computer science at the University of Southern California in 2011. His doctoral projects consisted of studying the data-driven representation learning algorithms and applying machine pipeline in industrial data analysis.
\end{IEEEbiography}
\vspace{-0.8cm}
\begin{IEEEbiography}[{\includegraphics[width=1in,height=1.25in,clip,keepaspectratio]{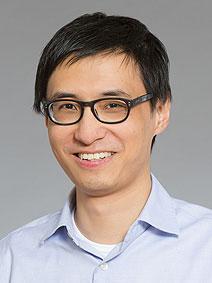}}]{Junming Yin}
is an assistant professor in the Department of Management Information Systems at the University of Arizona. Prior to that he was a Lane Fellow in the Lane Center of Computational Biology at Carnegie Mellon University. He received his Ph.D. from UC Berkeley in Computer Science. His research focus is on statistical machine learning and its applications in business intelligence, digital marketing, healthcare systems, and computational biology.
\end{IEEEbiography}
\vspace{-0.8cm}
\begin{IEEEbiography}[{\includegraphics[width=1in,height=1.25in,clip,keepaspectratio]{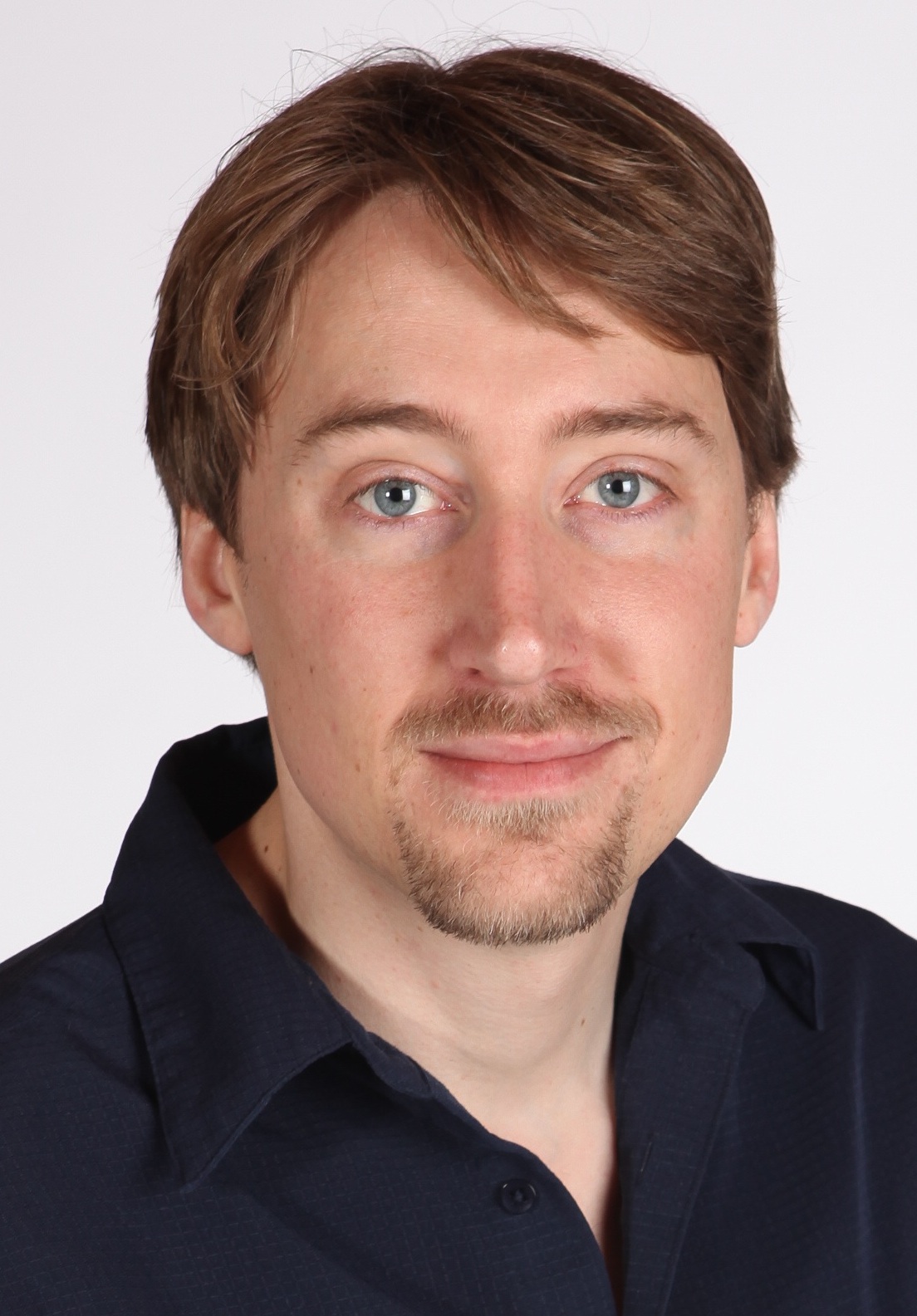}}]{Greg Ver Steeg}
is a research assistant professor in computer science at USC's Information Sciences Institute. He received his Ph.D. in physics from Caltech in 2009 and since then has focused on using ideas from information theory to understand complex systems like human behavior, biology, and language. His work has been recognized with an AFOSR Young Investigator Award.
\end{IEEEbiography}
\vspace{-0.8cm}
\begin{IEEEbiography}[{\includegraphics[width=1in,height=1.25in,clip,keepaspectratio]{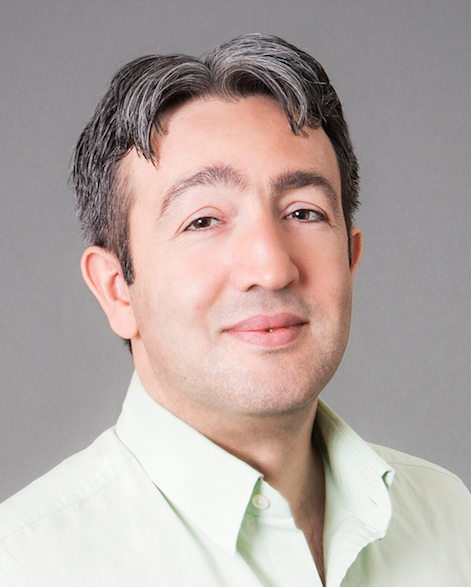}}]{Aram Galstyan}
is a research director for data science and machine learning at USC's Information Sciences Institute and a research associate professor in the USC Viterbi School of Engineering's Computer Science Department. His current research focuses on different aspects of machine learning, with applications in computational social sciences, forecasting, and bioinformatics.
\end{IEEEbiography}
\vspace{-0.8cm}
\newpage
\renewcommand{\thepage}{}
\setcounter{section}{0}
\renewcommand{\thesection}{\Alph{section}}
\twocolumn[{%
 \centering
 \LARGE
 Supplemental Materials to ``Scalable Temporal Latent Space Inference for Link Prediction in Dynamic Social Networks"\\
 \vspace{0.6cm}
}]

\section{Proof of Lemma~\ref{lemma:lipschitz}}\label{app:1}
According to Eq.~\ref{equ:subopt}, we can obtain the gradient of $J(Z_{\tau}(u))$ at time stamp $\tau$
\begin{equation}\label{equ:grad1}
\small{
\begin{aligned}
\nabla J(Z_{\tau}(u))=&-\lambda (Z_{\tau-1}(u)+Z_{\tau+1}(u))\\
&+2Z_{\tau}(u)\sum_{v\not\in N(u)}Z_{\tau}(v)^TZ_{\tau}(v)\\
&-2\sum_{v\in N(u)}(G_{\tau}(u,v)-Z_{\tau}(u)Z_{\tau}(v)^T)Z_{\tau}(v)
 \end{aligned}
 }
\end{equation}

We rewrite the term $\sum_{v\not\in N(u)}Z_{\tau}(v)^TZ_{\tau}(v)$ as:
\begin{equation}\label{equ:cache}
\small{
\begin{aligned}
=&\sum_{v\in V} Z_{\tau}(v)^TZ_{\tau}(v)-\sum_{v\in N(u)}Z_{\tau}(v)^TZ_{\tau}(v)-Z_{\tau}(u)^TZ_{\tau}(u)\\
=&Z_{\tau}^TZ_{\tau}-\sum_{v\in N(u)}Z_{\tau}(v)^TZ_{\tau}(v)-Z_{\tau}(u)^TZ_{\tau}(u)
\end{aligned}
}
\end{equation}
By combining Eq.~\ref{equ:grad1} and Eq.~\ref{equ:cache}, we have:
\begin{equation}
\small{
\begin{aligned}
\nabla J(Z_{\tau}(u))=&-\lambda (Z_{\tau-1}(u)+Z_{\tau+1}(u))+2Z_{\tau}(u)Z_{\tau}^TZ_{\tau}\\
&-2Z_{\tau}(u)Z_{\tau}(u)^TZ_{\tau}(u)-2\sum_{v\in N(u)}G_{\tau}(u,v)Z_{\tau}(v)
 \end{aligned}
}
\end{equation}

Since for each $u\in V$, $Z_{\tau}(u)Z_{\tau}(u)^T=1$, the gradient of $J(Z_{\tau}(u))$ can be simplified as
\begin{equation}\label{equ:particial}
\small{
\begin{aligned}
\nabla J(Z_{\tau}(u))=&-\lambda (Z_{\tau-1}(u)+Z_{\tau+1}(u))+2Z_{\tau}(u)Z_{\tau}^TZ_{\tau}\\
&-2Z_{\tau}(u)-2\sum_{v\in N(u)}G_{\tau}(u,v)Z_{\tau}(v)
 \end{aligned}
}
\end{equation}

For any two matrices $Z_1$, $Z_2$ $\in R^{1\times k}$, we replace $Z_{\tau}(u)$ in Eq.~\ref{equ:particial} with $Z_1$ and $Z_2$, we then have
\begin{equation*}
\small{
\begin{aligned}
&\|\nabla J(Z_1)-\nabla J(Z_2)\|_2^2\\
=&\|2(Z_1-Z_2)Z_{\tau}^TZ_{\tau}-(Z_1-Z_2)\|_2^2\\
=&\|2(Z_1-Z_2)(Z_{\tau}^TZ_{\tau}-I)\|_2^2, \\
=&4\texttt{tr}(((Z_1-Z_2)(Z_{\tau}^TZ_{\tau}-I))^T(Z_1-Z_2)(Z_{\tau}^TZ_{\tau}-I)), \\
=&4\texttt{tr}((Z_1-Z_2)^T(Z_1-Z_2)(Z_{\tau}^TZ_{\tau}-I)(Z_{\tau}^TZ_{\tau}-I)^T),
\end{aligned}
}
\end{equation*}
Since for any two positive semi-definite matrices $A$ and $B$, we have $\texttt{tr}(AB)\leq \texttt{tr}(A)\texttt{tr}(B)$. By substituting $A=(Z_1-Z_2)^T(Z_1-Z_2)$, $B=(Z_{\tau}^TZ_{\tau}-I)(Z_{\tau}^TZ_{\tau}-I)^T$, we have
\begin{equation*}
\small{
\begin{aligned}
&\|\nabla J(Z_1)-\nabla J(Z_2)\|_2^2\\
=&4\texttt{tr}((Z_1-Z_2)^T(Z_1-Z_2)(Z_{\tau}^TZ_{\tau}-I)(Z_{\tau}^TZ_{\tau}-I))^T), \\
\leq&4\texttt{tr}((Z_{\tau}^TZ_{\tau}-I)(Z_{\tau}^TZ_{\tau}-I)^T)\texttt{tr}((Z_1-Z_2)^T(Z_1-Z_2)), \\
=&4\texttt{tr}((Z_{\tau}^TZ_{\tau}-I)(Z_{\tau}^TZ_{\tau}-I)^T)\|Z_1-Z_2\|_2^2
\end{aligned}
}
\end{equation*}
Let us apply the above trace inequality again and using $\texttt{tr}(Z_{\tau}^TZ_{\tau})=n$, $\texttt{tr}(I)=k$, we then have
\begin{equation*}
\small{
\begin{aligned}
&\|\nabla J(Z_1)-\nabla J(Z_2)\|_2^2\\
\leq&4\texttt{tr}((Z_{\tau}^TZ_{\tau}-I)(Z_{\tau}^TZ_{\tau}-I)^T)\|Z_1-Z_2\|_2^2\\
\leq&4(n^2-2n+k)\|Z_1-Z_2\|_2^2
\end{aligned}
}
\end{equation*}
Thus, we have $\|\nabla J(Z_1)-\nabla J(Z_2)\|_2\leq 2\sqrt{(n^2-2n+k)}\|Z_1-Z_2\|_2$. Therefore, $\nabla J (Z_{\tau}(u))$ is Lipschitz continuous and the Lipschitz
constant is $2\sqrt{(n^2-2n+k)}$. This completes the proof.
\section{Proof of Lemma~\ref{lemma:updaterule}}\label{app:2}
Based on~\cite{opac-b1104789,journals/tsp/GuanTLY12,beck2009fast}, for each node $u$, at time point $\tau$, we could iteratively update $Z_{\tau}$(u) in each iteration $r+1$ with the following rule:
\begin{equation*}
\small{
\begin{aligned}
&Y^{(r+1)}=Y^{(r)}-\eta \nabla_{Z_{\tau}(u)}J(Y^{(r)})\\
&Z_{\tau}^{(r+1)}(u)=P_L(Y)=\max(Y^{(r+1)},0)\\
&Y^{(r+1)}=Z_{\tau}^{(r+1)}(u)+\frac{a_r-1}{a_{r+1}}(Z_{\tau}^{(r+1)}(u)-Z_{\tau}^{(r)}(u))
\end{aligned}
}
\end{equation*}
where $a_r$ is defined in Eq.~\ref{eq:ar}, $P_L(Y)=\argmin_{X\geq 0}\phi (Y,X)$ denotes the non-negative projection of $Y$, and $Y^{(0)}=Z_{\tau}^{I}(u)$ ($Z_{\tau}(u)^I$ is the initialization for $Z_{\tau}(u)$).

Let us replace the $\nabla_{Z_{\tau}(u)}J(Z_{\tau}^{(r)}(u))$ with Eq.~\ref{equ:particial} and the step size $\eta$ with $\frac{1}{L}$ ($L$ is the Lipschitz constant), which leads to the expression in Lemma~\ref{lemma:updaterule}.

Therefore, the update rule in Lemma~\ref{lemma:updaterule} is correct.

\section{Proof of Theorem~\ref{theory:errorrate}}\label{app:3}
According to Theorem 2.2.2 in~\cite{opac-b1104789} or Lemma 2.3~\cite{beck2009fast}, for any matrix $X\in R_{+}^{1\times k}$ and matrix $Y\in R^{1\times k}$, we have
\begin{equation}\label{equ:0}
\small{
J(X)\geq J(P_L(Y))+L<P_L(Y)-Y,Y-X>+\frac{L}{2}\|Y-P_L(Y)\|_F^2
}
\end{equation}
where $P_L(Y)=\argmin_{X\geq 0}\phi (Y,X)$ denotes the non-negative projection of $Y$, and $\langle, \rangle$ denotes the matrix inner product operator.

Now let $Y^{(r+1)}=Z^{(r)}(u)$ $+\frac{a_{r}-1}{a_{r+1}}$ $(Z^{(r)}(u)-Z^{(r-1)}(u))$ and $P_L(Y^{(r)})=Z^{(r)}(u)$. Note that we omitted the subscript $\tau$ that denotes the time to simplify notations. By substituting $X=Z^{(r)}(u)$, $Y=Y^{(r+1)}$, we have
\begin{equation}\label{equ:1}
\small{
\begin{aligned}
J(Z^{(r)}(u))\geq &J(Z^{(r+1)}(u))+\frac{L}{2}\|Y^{(r+1)}-Z^{(r+1)}(u)\|_F^2\\
&+L\langle Z^{(r+1)}(u)-Y^{(r+1)}(u), Y^{(r+1)}-Z^{(r)}(u)\rangle
\end{aligned}
}
\end{equation}
Similarly, by substituting $X=Z^*(u)$, $Y=Y^{(r+1)}$, we have
\begin{equation}\label{equ:2}
\small{
\begin{aligned}
J(Z^*(u))\geq&J(Z^{(r+1)}(u))+\frac{L}{2}\|Y^{(r+1)}-Z^{(r+1)}(u)\|_F^2\\
&+L \langle Z^{(r+1)}(u)-Y^{(r+1)}(u), Y^{(r+1)}-Z^*(u)\rangle
\end{aligned}
}
\end{equation}
since $a_{r+1}>1$, by multiplying both sides of Eq.~\ref{equ:1} by $a_{r+1}-1$ and adding it to Eq.~\ref{equ:2}, we have
\begin{equation}\label{equ:3}
\small{
\begin{aligned}
&(a_{r+1}-1)J(Z^{(r)}(u))+J(Z^*(u))\geq L \langle Z^{(r+1)}(u)-Y^{(r+1)},\\
&, a_{r+1}Y^{(r+1)}-(a_{r+1}-1)Z^{(r)}(u)-Z^*(u) \rangle\\
&+a_{r+1}J(Z^{(r+1)}(u))+\frac{La_{r+1}}{2}\|Y^{(r+1)}-Z^{(r+1)}(u)\|_F^2
\end{aligned}
}
\end{equation}
From Line 3 in Algorithm~\ref{alg:bcgd1}, we get $a_{r}^2= a_{r+1}^2-a_{r+1}$. By using this equality and multiplying both sides of Eq.~\ref{equ:3}, we have
\begin{equation}\label{equ:4}
\small{
\begin{aligned}
&a_r^2(J(Z^{(r)}(u))-J(Z^*(u)))-a_{r+1}^2(J(Z^{(r+1)}(u))-J(Z^*(u)))\\
&\geq \frac{L}{2} (\|a_{r+1}Z^{(r+1)}(u)-a_{r+1}Y^{(r+1)}\|_F^2+2\langle a_{r+1}Z^{(r+1)}(u)\\
&-a_{r+1}Y^{(r+1)}, a_{r+1}Y^{(r+1)}-(a_{r+1}-1)Z^{(r)}(u)-Z^*(u)\rangle)
\end{aligned}
}
\end{equation}
Since for any matrices $A$, $B$ and $C$, we have $\|A-B\|_F^2+\|B-C\|_F^2+2\langle A-B, B-C\rangle=\|A-B+B-C\|_F^2=\|A-C\|_F^2$.  That is, $\|A-B\|_F^2+2\langle A-B, B-C\rangle =\|A-C\|_F^2-\|B-C\|_F^2$. By replacing $A$ with $a_{r+1}Z^{(r+1)}(u)$, $B$ with $a_{r+1}Y^{(r+1)}$, $C$ with $(a_{r+1}-1)Z^{(r)}(u)+Z^*(u)$, we simplify Eq.~\ref{equ:4} as
\begin{equation}\label{equ:5}
\small{
\begin{aligned}
&a_r^2(J(Z^{(r)}(u))-J(Z^*(u)))-a_{r+1}^2(J(Z^{(r+1)}(u))-J(Z^*(u)))\\
&\geq \frac{L}{2}(\|a_{r+1}Z^{(r+1)}(u)-(a_{r+1}-1)Z^{(r)}(u)-Z^*(u)\|_F^2\\
&-\|a_{r+1}Y^{(r+1)}-(a_{r+1}-1)Z^{(r)}(u)-Z^*(u)\|_F^2)\\
&= \frac{L}{2}(\|a_{r+1}Z^{(r+1)}(u)-(a_{r+1}-1)Z^{(r)}(u)-Z^*(u)\|_F^2\\
&-\|a_rZ^{(r)}(u)-(a_r-1)Z^{(r-1)}(u)-Z^*(u)\|_F^2
\end{aligned}
}
\end{equation}
The last equality holds due to $Y^{(r+1)}=Z^{(r)}(u)$ $+\frac{a_{r}-1}{a_{r+1}}$ $(Z^{(r)}(u)-Z^{(r-1)}(u))$.

By varying $r$=0 to $r-1$ and summing up all these expressions in Eq.~\ref{equ:5}, we get
\begin{equation}\label{equ:6}
\small{
\begin{aligned}
&J(Z^{(0)}(u))-J(Z^*(u))-a_r^2(J(Z^{(r)}(u))-J(Z^*(u)))\\
\geq&-\frac{L}{2}\|Z^{(0)}(u)-Z^*(u)\|_F^2
\end{aligned}
}
\end{equation}
On another hand, we substitute $X$ and $Y$ in Eq.~\ref{equ:0} by $X=Z^*(u)$ and $Y=Y^{(0)}$, we have
\begin{equation}\label{equ:7}
\small{
\begin{aligned}
&J(Z^{(0)}(u))-J(Z^*(u))\leq\\
& \frac{L}{2}(\|Y^{(0)}-Z^*(u)\|_F^2-\|Z^{(0)}(u)-Z^*(u)\|_F^2)
\end{aligned}
}
\end{equation}

By combining Eq.~\ref{equ:6} and Eq.~\ref{equ:7}, we have
\begin{equation}\label{equ:8}
\small{
a_{r}^2(J(Z^{(r)}(u))-J(Z^*(u)))\leq \frac{L}{2}\|Y^{(0)}-Z^*(u)\|_F^2
}
\end{equation}

By substituting $Y^{(0)}$ as $Z^I(u)$ and using $a_r\geq$ $(1+r/2\sqrt{\alpha_0/L})$ $=(1+r/2)$ in Lemma 2.2.4 in~\cite{opac-b1104789}, we get:
\begin{equation*}
\small{
J(Z^{(r)}(u))-J(Z^*(u))\leq \frac{2L\|Z^I(u)-Z^*(u)\|_F^2}{(r+2)^2}
}
\end{equation*}
This completes the proof for Theorem.~\ref{theory:errorrate}.

\section{Extra Results based on $\texttt{AUC}_{\texttt{PR}}$}
In this set of experiments, we evaluate the link prediction accuracy according to the measurement of $\texttt{AUC}_{\texttt{PR}}$.
\begin{figure}[!t]
\centering
\label{fig:PR1}\includegraphics[width=0.9\columnwidth]{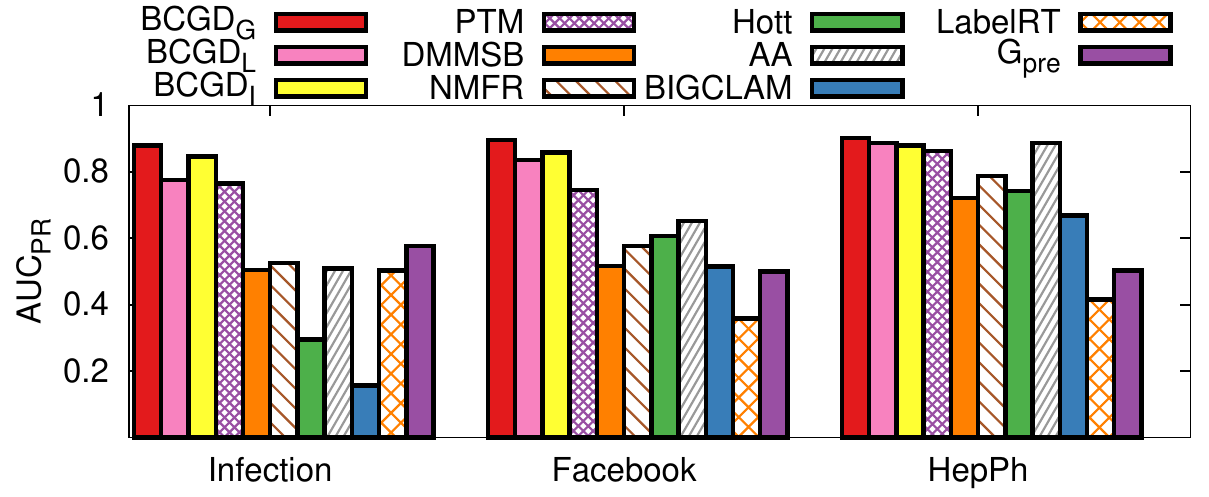}
\label{fig:PR2}\includegraphics[width=0.9\columnwidth]{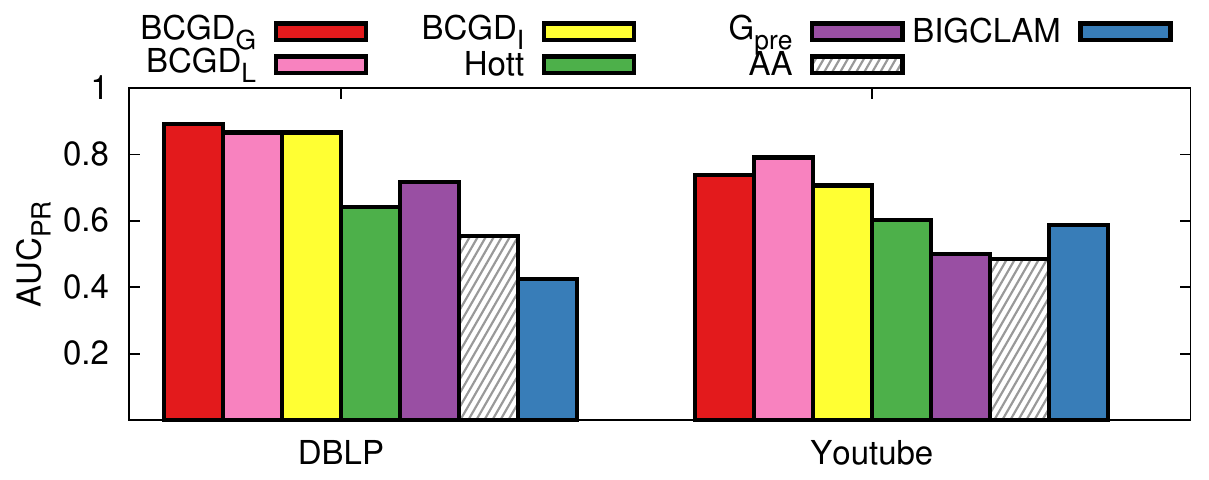}
\caption{\small{All links prediction accuracy comparison in terms of AUC under PR curves.}}\label{fig:PRall}
\vspace{-0.3cm}
\end{figure}

Figure~\ref{fig:PRall} gives the overall comparison of different approaches in terms of AUC under PR curves. The results
based on $\texttt{AUC}_{\texttt{PR}}$ are similar with those based on $\texttt{AUC}_{\texttt{ROC}}$: the proposed approaches are consistently better than other state-of-the-art approaches. We also observed that the performance of AA based on $\texttt{AUC}_{\texttt{PR}}$ is improved, possibly because AA gives high score for node pairs that have common neighbors and zero scores otherwise, which leads to high precision and relatively lower recall. To the contrary, the performance of DMMSB based on $\texttt{AUC}_{\texttt{PR}}$ is getting worse. All the results indicate that the proposed approach, especially the incremental algorithm $\texttt{BCGD}_I$, performs well in terms of link prediction task.
\end{document}